\documentclass[
 reprint,
 amsmath,
 amssymb,
 aps,
 prd,
 floatfix,
]{revtex4-2}

\usepackage{graphicx}% Include figure files
\usepackage{dcolumn}% Align table columns on decimal point
\usepackage{bm}% bold math
\usepackage{xcolor}
\colorlet{RED}{red}
\usepackage[normalem]{ulem}

\usepackage{tikz}
\usetikzlibrary{arrows.meta,decorations.pathmorphing}
\begin{document}

\preprint{APS/123-QED}

\title{High-Frequency Gravitational-Wave Transduction in a SQUID-Terminated Superconducting Cavity}% Force line breaks with \\

\author{H. Hadi}
\email{Hamedhadi1388@gmail.com}
\affiliation{Faculty of Physics, University of Tabriz, Tabriz, Iran}

\author{Amin Rezaei Akbarieh}
\email{amin.rezaeiakbarieh@kocaeli.edu.tr}
\affiliation{Faculty of Physics, University of Tabriz, Tabriz, Iran}
\affiliation{Department of Physics, Kocaeli University, 41001 Izmit, T\"urkiye}

\date{\today}% It is always \today, today,

\begin{abstract}
High-frequency gravitational waves in the MHz--GHz range require detection strategies beyond conventional interferometers. We study the response of a SQUID-terminated superconducting microwave cavity as a narrowband parametric transducer. The cavity boundary conditions and the flux dependence of the Josephson inductance are used to derive the quarter-wave mode spectrum and the first-order eigenfrequency response to changes in the physical line length, phase velocity, and SQUID inductive length. By projecting the perturbed dynamics onto the static cavity modes, we show that the resonance-frequency response and the photon-pair-production response are generally distinct. We therefore introduce two independent kernels, $R_n^{\omega}$ and $R_n^{\rm pair}$, which coincide only when changes in mode normalization and spatial profiles can be neglected. Near $\Omega_{\rm GW}\simeq 2\omega_n$, the isolated-mode dynamics reduce to a Mathieu-type parametric amplifier, allowing the photon number, quadrature variances, gain, and instability threshold to be obtained in the presence of dissipation. Spontaneous photon production scales quadratically with the gravitational-wave strain and is extremely small for representative parameters, whereas phase-sensitive responses with a coherent probe can scale linearly with strain. The framework therefore provides a theoretical description of narrowband gravitational-wave transduction, while a quantitative sensitivity estimate requires device-specific calibration of the mechanical--electromagnetic response, verification of mode isolation, and a complete treatment of loss and noise.

\end{abstract}

\maketitle

\section{Introduction}

LIGO and Virgo have provided direct detection of gravitational waves for studying  astrophysical and cosmological phenomena  \cite{LIGOScientific:2016aoc,LIGOScientific:2018mvr,KAGRA:2021vkt}. 
The current Kilometer-scale interferometers are not sensitive to frequencies higher than \(10\,{\rm Hz}\) to \(10^3\,{\rm Hz}\) \cite{Maggiore:2007ulw}. Therefore a large part of the possible gravitational-wave spectrum, particularly at much higher frequencies, remains outside their operational range.  Since some mechanisms associated with the early Universe, as well as certain nonstandard astrophysical sources, may generate gravitational waves in the MHz to GHz range \cite{Maggiore:1999vm,Aggarwal:2020olq}, finding a method to detect high gravitational frequencies waves is very important. Possible examples of such frequencies  include primordial backgrounds, cosmic-string networks, cosmological phase transitions, and ultralight-boson clouds around compact objects \cite{Caldwell:1991jj,Damour:2004kw,Caprini:2015zlo,Arvanitaki:2014wva,Brito:2017zvb}.

Some investigations are devoted to consider different approaches in recent years,  ranging from electromagnetic resonators and graviton--photon conversion to magnon excitations and radio observations in order to overcome the difficulty of detecting gravitational waves at such frequencies \cite{Cruise:2000za,Cruise:2005uq,Ejlli:2019bqj,Domcke:2020yzq,Ito:2022rxn,Aggarwal:2020olq}.

Among these approaches, superconducting microwave cavities have received particular attention \cite{Domcke:2023qle,Ballantini:2003nt,Berlin:2021txa,Berlin:2023grv,Fischer:2024msc}.  The main idea of  the early works based on coupled cavities, was that 
 gravitational-wave-induced deformations could transfer energy between electromagnetic modes. Following the early works, more recent studies have also considered the direct electromagnetic response of the cavity, mechanically mediated conversion, and superconducting radio-frequency systems \cite{Domcke:2023qle,Ballantini:2003nt,Berlin:2021txa,Berlin:2023grv,Fischer:2024msc}.   
 The important concept in such systems is distinguishing between   the direct effect of a gravitational wave on the electromagnetic field and the effect produced through mechanically induced changes in the cavity geometry. Note that  the strength and structure of the coupling ultimately depend on the detector-frame electromagnetic response, the spatial overlap of the modes, and the mechanical boundary conditions of the device \cite{Berlin:2021txa,Berlin:2023grv}.\\
 
Superconducting quantum circuits,  provide a suitable platform for studying quantum fields in engineered microwave environments \cite{Johansson:2009zz,Wilson:2011rsw,Nation:2011dka}.  The possibility of this platform is because of  the high tunability of their electromagnetic parameters.  As an important example in this context one can name the  dynamical Casimir effect, in which time-dependent changes in the boundary conditions can convert vacuum fluctuations into correlated photons \cite{Ford:1982ct,Dodonov:2010zza}.  This mechanism can be implemented in a controlled manner in a coplanar waveguide terminated by a superconducting quantum interference device (SQUID).       Its control manner is possible  because the Josephson inductance of the SQUID depends on the external magnetic flux, and the SQUID therefore acts as a tunable electromagnetic boundary.

According to above explanations, one can modulate the external flux rapidly to vary the effective electrical length of the resonator without mechanically displacing a physical boundary, thereby enabling various parametric processes, including degenerate and nondegenerate processes as well as intermode conversion \cite{Wustmann:2013icg,Bengtsson:2018vvv}.   In the geometry of the cavity that we consider in this paper, 
the series coupling capacitor produces an approximately open boundary at the readout end. For the opposite end, the SQUID provides a tunable Robin boundary ; consequently, the natural mode structure of the cavity is of the quarter-wave type which derived explicitly in Sec.~\ref{sec:squid_cavity}.  Such a configuration also makes it possible to investigate whether a gravitational wave can induce, through a parametric perturbation, dynamics in the cavity analogous to the dynamical Casimir effect \cite{deOliveira:2026pmn}. 

Notice that, the effect of  a gravitational wave cannot simply be reduced to a universal geometric change in the cavity length in a realistic detector. The geometric change in the cavity length can be proportional to  $hL_0/2$.   The response of the device to the gravitational wave depends on feature such as  its geometry, the electromagnetic properties of the circuit, the mechanical structure of its support, and the characteristics of the cavity modes.  Because of these properties, in this paper, the effect of the gravitational perturbation on the distributed circuit and its mechanical structure is described using transfer functions defined in the detector frame.   By projecting the perturbed quadratic action onto the static cavity modes, it becomes clear that the response of the system must be described by two independent kernels: $R_n^{\omega}$, which determines the measurable modulation of the resonance frequency, and $R_n^{\rm pair}$, which controls the strength of the resonant parametric interaction $a_n^{\dagger 2}+a_n^2$ and, consequently, the photon-pair-production process.  
This distinct feature  is crucial  because, if the gravitational wave modifies the resonance frequency and also the effective modal capacitance, impedance, normalization, or spatial mode structure, measurement of the frequency response alone is insufficient to determine the pair-production response. Only within the more restricted approximation in which the perturbation is treated solely as a frequency modulation and changes in the mode normalization and mode structure are neglected do the two kernels become equal, allowing the system dynamics to be described using the closed-form Mathieu formulation.

Distinguishing between device-specific transfer functions and universal parametric dynamics is fundamental. Accurate prediction of fabricated chip behavior requires knowledge of the elastic displacement field, anchoring and package boundary conditions, detector-frame changes in electromagnetic parameters, and the resulting mode-overlap integrals. These factors are not represented in a simple one-dimensional circuit schematic. Consequently, all observable scalings are expressed using $R_n^{\omega}$ and $R_n^{\rm pair}$, with values determined by a calibrated mechanical--electromagnetic model or independent modulation measurements.

According to the circuit Lagrangian, we first derive the resonator boundary conditions along with the quarter-wave spectrum. We then obtain an exact first-order formula for the variation of an eigenfrequency under changes of the physical length, phase velocity, and SQUID inductive length. A fixed-mode quadratic projection identifies the diagonal and intermode parametric couplings. The usual single-mode degenerate-parametric-amplifier description is maintained only when all unwanted sum- and difference-frequency channels are spectrally resolved. Finally we  include damping and input--output coupling, derive positivity-preserving moment equations, and compare spontaneous emission, coherent probing, and the flux-pumped response in its quasistatic and frequency-selective regimes.

The structure of this paper is as follows. Section~\ref{sec:squid_cavity} considers the capacitively coupled SQUID-cavity spectrum and its parameter sensitivities. Section~\ref{sec:gw_frequency_modulation} indicates the details of the detector-frame response kernels, the fixed-mode projection, and the criteria for single-mode validity. Section~\ref{sec:mode_equation_mathieu} illustrates  the restricted Mathieu approach. Section~\ref{sec:quantum_hamiltonian_photon_pair} presents the quantum Hamiltonian. Section~\ref{sec:dissipation_quality_sensitivity} analyzes dissipation and gain scaling. Detection channels and experimental considerations are investigated in Secs.~\ref{sec:detection_channels} and \ref{sec:detector_response}. Section~\ref{sec:numerical-simulation} discusses Gaussian dynamics as well as consistency verifications. Finally, Sec.~\ref{sec:discussion_conclusion} summarizes the main conclusions and limitations.

\section{SQUID-Terminated Coplanar Waveguide Cavity}
\label{sec:squid_cavity}

We consider a one-dimensional superconducting coplanar-waveguide cavity of physical length $L_0$, oriented along the unit vector $\mathbf n$. Its electromagnetic degree of freedom is the node-flux field $\phi(x,t)$, with $x\in[0,L_0]$. Away from the terminations, the linearized transmission-line Lagrangian is
\begin{equation}
\label{eq:tl_lagrangian}
\mathcal L_{\rm TL}
=
\int_0^{L_0}dx
\left[
\frac{C_\ell}{2}\dot\phi^2
-
\frac{1}{2L_\ell}(\partial_x\phi)^2
\right],
\end{equation}
where $C_\ell$, and $L_\ell$ denote the capacitance and inductance per unit length, respectively. Variation of this Lagrangian with respect to $\phi(x,t)$ yields
\begin{equation}
\label{eq:wave_equation}
    \frac{\partial^2\phi}{\partial t^2}
    -v^2\frac{\partial^2\phi}{\partial x^2}=0,
\end{equation}
with
\begin{equation}
\label{eq:phase_velocity}
    v\equiv\frac{1}{\sqrt{L_\ell C_\ell}}.
\end{equation}

The left end is coupled through a small series capacitance $C_c$ to a semi-infinite readout line. The corresponding boundary contribution is $C_c[\dot\phi(0,t)-\dot\phi_{\rm line}(0,t)]^2/2$, where the subscript ``line'' denotes the node-flux field of the semi-infinite readout transmission line evaluated at the coupling point. The same variation with respect to the cavity node-flux field
$\phi(x,t)$ that yields Eq.~\eqref{eq:wave_equation} in the bulk,
when applied to the total action including the coupling capacitor,
gives the following boundary condition at $x=0$:
\begin{equation}
\label{eq:coupling_capacitor_boundary}
\frac{1}{L_\ell}\partial_x\phi(0,t)
=
C_c\left[
\ddot\phi(0,t)-\ddot\phi_{\rm line}(0,t)
\right].
\end{equation}
For the closed-cavity eigenproblem, the weak-coupling condition $\omega C_c Z_0\ll1$, where $Z_0$ is the characteristic impedance
of the semi-infinite readout line and $\omega$ is the angular frequency of the cavity mode under consideration, reduces
Eq.~\eqref{eq:coupling_capacitor_boundary} to the open-end condition
\begin{equation}
\label{eq:fixed_boundary}
\partial_x\phi(0,t)=0.
\end{equation}
The finite coupling is reintroduced at the input--output level through the external decay rate $\kappa_{\rm ext}$, where the subscript ``ext'' denotes loss through coupling to the readout line. The total loaded linewidth is given by

\begin{equation}
\label{eq:total_kappa_internal_external}
    \kappa=\kappa_{\rm int}+\kappa_{\rm ext},
\end{equation}
where $\kappa_{\rm int}$ denotes the internal decay rate arising from intrinsic cavity losses. The termination at $x=L_0$ consists of a symmetric SQUID composed of two identical Josephson junctions. Away from the half-flux bias points, the effective Josephson energy is
\begin{equation}
\label{eq:squid_josephson_energy_flux}
    E_J(\Phi_{\rm ext})
    =
    2E_{J0}
    \left|
    \cos\left(\frac{\pi\Phi_{\rm ext}}{\Phi_0}\right)
    \right|,
\end{equation}
where $\Phi_{\rm ext}$ represents the external magnetic flux threading the SQUID loop, and $E_{J0}$ denotes the Josephson energy of each individual junction.
The subscript "0" designates the Josephson energy of a single junction, whereas the total effective Josephson energy at zero external flux is $2E_{J0}$. The energy of a single junction relates to the junction critical current by
\begin{equation}
\label{eq:josephson_energy_critical_current}
    E_{J0}
    =
    \frac{\Phi_0 I_c}{2\pi},
    \qquad
    \Phi_0
    =
    \frac{h}{2e},
\end{equation}
where $I_c$ denotes the critical current of each junction, $\Phi_0$ is the superconducting flux quantum, $h$ is Planck's constant, and $e$ is the elementary charge.
The linearized Josephson inductance is expressed as
\begin{equation}
\label{eq:josephson_inductance}
    L_J(\Phi_{\rm ext})
    =
    \left(\frac{\Phi_0}{2\pi}\right)^2
    \frac{1}{E_J(\Phi_{\rm ext})}.
\end{equation}
Including the effective junction capacitance $C_J$, the variation of the linearized SQUID action yields
\begin{equation}
\label{eq:full_squid_boundary}
\frac{1}{L_\ell}\partial_x\phi(L_0,t)
+
C_J\ddot\phi(L_0,t)
+
\frac{\phi(L_0,t)}{L_J}=0.
\end{equation}
The capacitance of the SQUID can be neglected provided that every relevant angular frequencies are significantly lower than the SQUID plasma frequency:
\begin{equation}
\label{eq:squid_plasma_frequency_condition}
    \omega_{n0},\,
    \Omega_{\rm drive},\,
    \Omega_{\rm GW}
    \ll\omega_p,
\end{equation}
where $\omega_{n0}$ denotes the unperturbed angular eigenfrequency of the $n$th cavity mode. $\Omega_{\rm drive}$ denotes the angular frequency of
the externally applied circuit drive, while $\Omega_{\rm GW}$ denotes the angular frequency of the incident gravitational wave. The SQUID
plasma frequency is defined as
\begin{equation}
\label{eq:squid_plasma_frequency}
    \omega_p=\frac{1}{\sqrt{L_JC_J}},
\end{equation}
where $L_J$ and $C_J$ denote the effective Josephson inductance and capacitance of the SQUID termination. Under these assumptions, Equation~\eqref{eq:full_squid_boundary} reduces to 
\begin{equation}
\label{eq:robin_boundary}
    \phi(L_0,t)
    +L_{\rm SQ}(t)\,
    \partial_x\phi(L_0,t)=0,
\end{equation}
with
\begin{equation}
\label{eq:lsq_lj_relation}
    L_{\rm SQ}(t)=\frac{L_J(t)}{L_\ell}.
\end{equation}
therefore,
\begin{equation}
\label{eq:lsq_flux}
    L_{\rm SQ}(\Phi_{\rm ext})
    =
    \frac{1}{L_\ell}
    \left(\frac{\Phi_0}{2\pi}\right)^2
    \frac{1}{E_J(\Phi_{\rm ext})}.
\end{equation}
For a static bias, the time dependence of the field can be separated from its spatial dependence by writing $\phi(x,t)=u_n(x)e^{-i\omega_nt}$, which, upon substitution into the wave equation, leads to the following equation for the spatial mode function:
\begin{equation}
\label{eq:spatial_eigenvalue_equation}
    \frac{d^2u_n}{dx^2}+k_n^2u_n=0,
    \qquad
    \omega_n=vk_n.
\end{equation}
The form of the mode function is then fixed by the boundary condition at the open end of the cavity, and imposing Eq.~\eqref{eq:fixed_boundary} gives
\begin{equation}
\label{eq:mode_function}
    u_n(x)=A_n\cos(k_nx).
\end{equation}
Applying the SQUID boundary condition in Eq.~\eqref{eq:robin_boundary} to the mode function gives the following condition for the allowed values of $k_n$:
\begin{equation}
\label{eq:quantization_condition_cos}
    \cos(k_nL_0)
    -k_nL_{\rm SQ}\sin(k_nL_0)=0,
\end{equation}
which can be written equivalently as
\begin{equation}
\label{eq:quantization_condition_tan}
    \cot(k_nL_0)=k_nL_{\rm SQ},
    \qquad
    \tan(k_nL_0)=\frac{1}{k_nL_{\rm SQ}}.
\end{equation}
These relations determine the allowed cavity modes and lead to the quarter-wave spectrum associated with the capacitively coupled geometry shown in Fig.~\ref{fig:squid_gw_detector}.

In the limit $k_nL_{\rm SQ}\ll1$, the allowed wavenumbers are expected to lie close to the quarter-wave values, so we write $k_nL_0=(n-\tfrac12)\pi-\delta_n$, where $\delta_n$ is a small correction. Using $\cot[(n-\tfrac12)\pi-\delta_n]\simeq\delta_n$ in Eq.~\eqref{eq:quantization_condition_tan}, we find $\delta_n\simeq k_nL_{\rm SQ}$, which leads to the approximate wavenumbers
\begin{equation}
\label{eq:approx_wavenumber}
    k_{n0}
    \simeq
    \frac{(n-\tfrac12)\pi}
    {L_0+L_{\rm SQ}^{(0)}},
    \qquad n=1,2,3,\ldots,
\end{equation}
and, using $\omega_{n0}=vk_{n0}$, the corresponding unperturbed mode frequencies are
\begin{equation}
\label{eq:unperturbed_frequencies}
    \omega_{n0}
    \simeq
    \frac{(n-\tfrac12)\pi v}
    {L_0+L_{\rm SQ}^{(0)}}.
\end{equation}
Here,
\begin{equation}
    L_{\rm SQ}^{(0)}
    \equiv
    \frac{L_J(\Phi_b)}{L_\ell}
\end{equation}
denotes the static, unperturbed inductive length of the SQUID at the chosen external flux bias $\Phi_b$. For multimode resonance criteria, the approximate harmonic spectrum is not sufficient, so the exact roots of Eq.~\eqref{eq:quantization_condition_tan} should be used. The corresponding first-order variation of the eigenvalue follows by differentiating
\begin{equation}
\label{eq:eigenvalue_function}
F_n(k_n,L_0,L_{\rm SQ})
\equiv
\cos(k_nL_0)-k_nL_{\rm SQ}\sin(k_nL_0)=0.
\end{equation}
Using $\cos(k_nL_0)=k_nL_{\rm SQ}\sin(k_nL_0)$ in the resulting expression, we obtain
\begin{equation}
\label{eq:exact_k_variation}
\frac{\delta k_n}{k_n}
=
-
\frac{
[1+(k_nL_{\rm SQ})^2]\,\delta L_0
+
\delta L_{\rm SQ}
}
{
L_0[1+(k_nL_{\rm SQ})^2]+L_{\rm SQ}
}.
\end{equation}
To write this result more compactly, we introduce the dimensionless participation factors
\begin{align}
\label{eq:length_participations}
P_{L,n}
&=
\frac{L_0[1+(k_nL_{\rm SQ})^2]}
{L_0[1+(k_nL_{\rm SQ})^2]+L_{\rm SQ}},
\\
P_{{\rm SQ},n}
&=
\frac{L_{\rm SQ}}
{L_0[1+(k_nL_{\rm SQ})^2]+L_{\rm SQ}},
\end{align}
which satisfy $P_{L,n}+P_{{\rm SQ},n}=1$. Since $\omega_n=vk_n$, the exact diagonal frequency variation within the one-dimensional boundary model can then be written as
\begin{equation}
\label{eq:exact_frequency_variation}
\frac{\delta\omega_n}{\omega_n}
=
\frac{\delta v}{v}
-
P_{L,n}\frac{\delta L_0}{L_0}
-
P_{{\rm SQ},n}\frac{\delta L_{\rm SQ}}{L_{\rm SQ}}.
\end{equation}
In the limit $k_nL_{\rm SQ}\ll1$, we have $P_{L,n}\simeq L_0/(L_0+L_{\rm SQ})$, so the familiar dilution factor appears as an intrinsic part of the response rather than as an additional multiplicative factor.

Direct external-flux modulation and gravitational transduction are physically distinct mechanisms, so their contributions are separated as
\begin{equation}
\label{eq:parameter_decomposition}
\delta X(t)
=
\delta X^{\rm drive}(t)
+
\delta X^{\rm GW}(t),
\qquad
X\in\{v,L_0,L_{\rm SQ}\}.
\end{equation}
Here, the first term describes a controlled circuit drive, whereas the second must be obtained from the mechanical and electromagnetic response in the detector frame. Once these parameter perturbations are identified, Eq.~\eqref{eq:exact_frequency_variation} gives the corresponding eigenfrequency sensitivity, but it does not determine the perturbations themselves.

\section{Operational Gravitational-Wave Response and Modal Projection}
\label{sec:gw_frequency_modulation}

\subsection{Detector-frame transfer functions}

We define the gravitational response operationally by separating the exact circuit sensitivity derived in Sec.~\ref{sec:squid_cavity} from the device-specific problem of mapping a detector-frame metric perturbation onto changes in the line parameters, substrate displacement field, SQUID environment, and electromagnetic constitutive tensors. For a monochromatic plane wave, this perturbation is written as

\begin{align}
\label{eq:gw_plane_wave}
h_{ij}(t,\mathbf x)
&=
\operatorname{Re}\!\left[
h_0e_{ij}(\hat{\mathbf k},\psi)
 e^{-i\Omega_{\rm GW}t+i\mathbf k_{\rm GW}\cdot\mathbf x}
\right],
\\
\mathbf k_{\rm GW}
&=
\frac{\Omega_{\rm GW}}{c}\hat{\mathbf k}.
\end{align}
For a device of size $L_{\rm det}$, the long-wavelength expansion is valid when $k_{\rm GW}L_{\rm det}\ll1$, whereas phase averaging along a straight line segment is controlled by $|\mathbf k_{\rm GW}\cdot\mathbf n|L_0$.

In the standard transverse-traceless (TT) gauge, the response of an ideal, freely responding line element is given by
\begin{equation}
\label{eq:ideal_free_mass_length_response}
    \frac{\delta L(t)}{L_0}
    =
    \frac{1}{2L_0}
    \int_{-L_0/2}^{L_0/2}
    n^in^jh_{ij}\bigl(t,\mathbf x_c+\mathbf ns\bigr)\,ds.
\end{equation}
In the long-wavelength limit, this reduces to $\delta L/L_0=n^in^jh_{ij}(t,\mathbf x_c)/2$. Equation~\eqref{eq:ideal_free_mass_length_response}, however, serves only as an idealized reference and does not describe the strain experienced by a chip supported by a substrate. The response of the actual device is instead represented by the dimensionless complex transfer functions

\begin{align}
\label{eq:device_transfer_functions}
\frac{\delta v^{\rm GW}(t)}{v}
&=
\operatorname{Re}\!\left[
T_{v,n}(\Omega_{\rm GW})h_0e^{-i\Omega_{\rm GW}t}
\right],
\\
\frac{\delta L_0^{\rm GW}(t)}{L_0}
&=
\operatorname{Re}\!\left[
T_{L,n}(\Omega_{\rm GW})h_0e^{-i\Omega_{\rm GW}t}
\right],
\\
\frac{\delta L_{\rm SQ}^{\rm GW}(t)}{L_{\rm SQ}}
&=
\operatorname{Re}\!\left[
T_{{\rm SQ},n}(\Omega_{\rm GW})h_0e^{-i\Omega_{\rm GW}t}
\right].
\end{align}
Here, $T_{v,n}$, $T_{L,n}$, and $T_{{\rm SQ},n}$ describe, respectively, the gravitational-wave-induced fractional responses of the phase velocity, the physical line length, and the SQUID inductive length. Their magnitudes give the response amplitudes per unit gravitational-wave strain, while their arguments specify the phase shifts relative to the incident wave. The first subscript identifies the affected device parameter, and $n$ denotes the cavity-mode index. Any direct detector-frame electromagnetic contribution that cannot be expressed through these three effective parameters is denoted by $R_n^{\rm dir}$. Combining Eqs.~\eqref{eq:exact_frequency_variation} and \eqref{eq:device_transfer_functions}, we define the ordinary frequency-response function as
\begin{equation}
\label{eq:frequency_response_function}
R_n^{\omega}
=
T_{v,n}
-
P_{L,n}T_{L,n}
-
P_{{\rm SQ},n}T_{{\rm SQ},n}
+
R_n^{\rm dir},
\end{equation}
so that
\begin{equation}
\label{eq:frequency_response_definition}
\frac{\delta\omega_n^{\rm GW}(t)}{\omega_{n0}}
=
\operatorname{Re}\!\left[
R_n^{\omega}
 h_0e^{-i\Omega_{\rm GW}t}
\right].
\end{equation}
The response $R_n^{\omega}$ already contains the effects of geometric dilution, polarization, incidence angle, finite wavelength, and mechanical and electromagnetic transfer. Therefore, no additional length-dilution factor should be introduced in the subsequent formulas.

For the ideal reference model, in which only the physical line length responds geometrically while $v$ and $L_{\rm SQ}$ remain fixed, the response reduces to
\begin{equation}
\label{eq:ideal_response_limit_with_squid_dilution}
R_n^{\omega}
\rightarrow
-
\frac{P_{L,n}}{2}
 n^in^je_{ij}(\hat{\mathbf k},\psi)
\mathcal A_n(\mathbf k_{\rm GW}),
\end{equation}
where the spatial averaging factor is defined by
\begin{equation}
\label{eq:finite_wavelength_response}
\mathcal A_n(\mathbf k_{\rm GW})
=
\int_0^{L_0} dx\,W_n(x)
e^{i\mathbf k_{\rm GW}\cdot\mathbf x(x)},
\end{equation}
and the weighting function is normalized according to
\begin{equation}
\label{eq:weight_function_normalization}
\int_0^{L_0} W_n(x)\,dx
=
1.
\end{equation}
For uniform weighting along a straight segment centered at $\mathbf x_c$, this factor becomes
\begin{equation}
\label{eq:finite_wavelength_overlap_sinc}
\mathcal A_n
=
 e^{i\mathbf k_{\rm GW}\cdot\mathbf x_c}
\operatorname{sinc}\!\left(
\frac{\mathbf k_{\rm GW}\cdot\mathbf n\,L_0}{2}
\right),
\qquad
\operatorname{sinc}x=\frac{\sin x}{x}.
\end{equation}
For a fabricated device, both $W_n$ and the transfer functions in Eq.~\eqref{eq:device_transfer_functions} must be obtained from the actual elastic and electromagnetic problem.

Introducing
\begin{equation}
\label{eq:eta_omega_definition}
\eta_n^{\omega}
\equiv
2|R_n^{\omega}|,
\qquad
\theta_n^{\omega}=\arg R_n^{\omega},
\end{equation}
the amplitude of the resonance-frequency modulation can be written as
\begin{equation}
\label{eq:gw_frequency_shift_amplitude}
\delta\omega_{{\rm GW},n}
=
\omega_{n0}|R_n^{\omega}|h_0
=
\frac{\eta_n^{\omega}h_0\omega_{n0}}{2}.
\end{equation}
The transfer factor $\eta_n^{\omega}$ is not generally bounded by unity, since the response may be suppressed, vanish, acquire a phase, or be resonantly enhanced by the mechanical response.

\subsection{Fixed-mode quadratic action and the pair-response kernel}

A time-dependent eigenfrequency alone does not specify the quantum pair-production Hamiltonian if the perturbation also changes the modal capacitance or normalization. To make this distinction explicit, expand the flux field in the static, capacitively normalized modes, $\phi(x,t)=\sum_nq_n(t)u_n(x)$. After neglecting $C_J$ under the frequency condition in Eq.~\eqref{eq:squid_plasma_frequency_condition}, the static modes are normalized according to
\begin{equation}
\label{eq:static_mode_normalization}
\int_0^{L_0}dx\,C_\ell u_m(x)u_n(x)
=
\delta_{mn}.
\end{equation}
For a reciprocal transmission-line perturbation, and after removing total derivatives, the minimal quadratic modal Lagrangian may be written as
\begin{equation}
\label{eq:general_modal_lagrangian}
\begin{split}
\mathcal L
={}&
\frac12\sum_n
\left(\dot q_n^2-\omega_{n0}^2q_n^2\right)
\\
&+
\frac12\sum_{mn}
\left[
C_{mn}(t)\dot q_m\dot q_n
-
K_{mn}(t)q_mq_n
\right],
\end{split}
\end{equation}
where $C_{mn}$ is dimensionless and $K_{mn}$ has dimensions of frequency squared. The matrices are obtained from overlap integrals between the detector-frame perturbation and the unperturbed mode profiles. For fixed coordinate boundaries, perturbations of the distributed capacitance, inverse inductance, and SQUID inverse inductance give
\begin{align}
\label{eq:explicit_modal_overlap_matrices}
C_{mn}(t)
&=
\int_0^{L_0}dx\,
\delta C_\ell(x,t)u_m(x)u_n(x),
\\
K_{mn}(t)
&=
\int_0^{L_0}dx\,
\delta\!\left[L_\ell^{-1}(x,t)\right]
\partial_xu_m(x)\partial_xu_n(x)
\nonumber\\
&\quad+
\delta\!\left[L_J^{-1}(t)\right]
u_m(L_0)u_n(L_0).
\end{align}
A moving material boundary, substrate strain, or detector-frame perturbation of the electromagnetic constitutive tensors adds the corresponding geometric and field-overlap terms to Eq.~\eqref{eq:explicit_modal_overlap_matrices}. For notational simplicity, we have omitted a possible mixed term $\dot{\mathbf q}^{\mathsf T}\mathbf A(t)\mathbf q$. Its symmetric part can be absorbed into $\mathbf K$ through integration by parts, whereas an antisymmetric part produces additional off-diagonal couplings and must be retained in a complete detector calculation. Once the removable part is absorbed into $K_{nn}$, this term does not affect the diagonal single-mode distinction derived below. To first order, the canonical Hamiltonian takes the form
\begin{equation}
\label{eq:general_modal_hamiltonian}
\begin{split}
H
={}&
\frac12\sum_n
\left(p_n^2+\omega_{n0}^2q_n^2\right)
\\
&-
\frac12\sum_{mn}C_{mn}(t)p_mp_n
+
\frac12\sum_{mn}K_{mn}(t)q_mq_n.
\end{split}
\end{equation}
For the diagonal element of mode $n$, we write
\begin{align}
\label{eq:CK_complex_amplitudes}
C_{nn}(t)
&=
\operatorname{Re}\!\left[
\mathcal C_n(\Omega_{\rm GW})
h_0e^{-i\Omega_{\rm GW}t}
\right],
\\
\frac{K_{nn}(t)}{\omega_{n0}^2}
&=
\operatorname{Re}\!\left[
\mathcal K_n(\Omega_{\rm GW})
h_0e^{-i\Omega_{\rm GW}t}
\right].
\end{align}
The dimensionless complex amplitudes $\mathcal C_n$ and $\mathcal K_n$ describe, respectively, the diagonal kinetic and stiffness perturbations of mode $n$. In particular, $\mathcal C_n$ represents the gravitational-wave-induced modulation of the modal kinetic normalization, or equivalently the effective modal capacitance, while $\mathcal K_n$ represents the corresponding modulation of the modal stiffness normalized by $\omega_{n0}^2$. Their magnitudes determine the modulation amplitudes per unit gravitational-wave strain, and their arguments give the phase shifts relative to the incident wave. A static perturbation changes the eigenfrequency according to
\begin{equation}
\label{eq:Romega_from_CK}
R_n^{\omega}
=
\frac12\left(\mathcal K_n-\mathcal C_n\right),
\end{equation}
whereas the coefficient of $a_n^{\dagger2}+a_n^2$ is determined by
\begin{equation}
\label{eq:Rpair_definition}
R_n^{\rm pair}
\equiv
\frac12\left(\mathcal K_n+\mathcal C_n\right).
\end{equation}
Therefore, $R_n^{\rm pair}=R_n^{\omega}$ only when $\mathcal C_n\simeq0$, meaning that the perturbation can be treated as a pure stiffness or frequency modulation with a fixed modal normalization. We define
\begin{equation}
\label{eq:eta_pair_definition}
\eta_n^{\rm pair}
\equiv
2|R_n^{\rm pair}|.
\end{equation}
The spontaneous-pair and squeezing estimates below depend on $\eta_n^{\rm pair}$, while the dispersive coherent sidebands depend on $\eta_n^{\omega}$. In the pure-stiffness approximation for the ideal freely responding line in Eq.~\eqref{eq:ideal_response_limit_with_squid_dilution}, $\mathcal C_n=0$, and Eqs.~\eqref{eq:Romega_from_CK} and \eqref{eq:Rpair_definition} give
\begin{align}
\label{eq:ideal_pair_and_frequency_kernels}
R_{n,{\rm ref}}^{\rm pair}
&=
R_{n,{\rm ref}}^{\omega}
=
-\frac{P_{L,n}}{2}
 n^in^je_{ij}(\hat{\mathbf k},\psi)
\mathcal A_n(\mathbf k_{\rm GW}),
\\
\eta_{n,{\rm ref}}^{\rm pair}
&=
\eta_{n,{\rm ref}}^{\omega}
=
P_{L,n}
\left|
 n^in^je_{ij}(\hat{\mathbf k},\psi)
\mathcal A_n(\mathbf k_{\rm GW})
\right|.
\end{align}
This reference response is nonzero whenever both the polarization projection and the spatial overlap are nonzero. It is used below only as an analytical benchmark, since the response of a supported chip must be obtained from the complete detector-frame overlaps.

\subsection{Multimode structure and validity of the single-mode reduction}

The off-diagonal elements of Eq.~\eqref{eq:general_modal_hamiltonian} give rise to both pair-creation and mode-conversion processes. In the interaction picture, their resonant contribution can be written schematically as
\begin{equation}
\label{eq:multimode_rwa_hamiltonian}
\begin{split}
H_{\rm int}^{\rm multi}
={}&
 i\hbar\sum_{m\le l}
\left[
\lambda_{ml}a_m^\dagger a_l^\dagger
-\lambda_{ml}^*a_ma_l
\right]
\\
&+
\hbar\sum_{m\ne l}
\left[
g_{ml}a_m^\dagger a_l+g_{ml}^*a_l^\dagger a_m
\right].
\end{split}
\end{equation}
The complex coefficients $\lambda_{ml}$ and $g_{ml}$ are determined by the off-diagonal modal overlaps: $\lambda_{ml}$ describes pair creation or annihilation between modes $m$ and $l$, while $g_{ml}$ describes frequency conversion between them. Accordingly, the first sum becomes resonant near $\Omega_{\rm GW}=\omega_m+\omega_l$, whereas the second becomes resonant near $\Omega_{\rm GW}=|\omega_m-\omega_l|$. A single-mode description of mode $n$ is therefore valid only if every unwanted pair $(m,l)$ satisfies
\begin{equation}
\label{eq:single_mode_isolation_condition}
\begin{split}
\left|\Omega_{\rm GW}-(\omega_m+\omega_l)\right|
&\gg
\max\{\kappa_m,\kappa_l,|\lambda_{ml}|\},
\\
\left|\Omega_{\rm GW}-|\omega_m-\omega_l|\right|
&\gg
\max\{\kappa_m,\kappa_l,|g_{ml}|\}.
\end{split}
\end{equation}
Since the quarter-wave spectrum is approximately harmonic, these isolation conditions are not automatically satisfied. Indeed, using Eq.~\eqref{eq:unperturbed_frequencies}, with $\omega_j\simeq(2j-1)\omega_1$, gives
\begin{align}
\label{eq:harmonic_multimode_degeneracies}
\omega_{m+2n-1}-\omega_m
&\simeq
2\omega_n,
\qquad m\geq1,
\\
\omega_m+\omega_{2n-m}
&\simeq
2\omega_n,
\qquad
1\leq m\leq2n-1,
\quad m\neq n.
\end{align}
The first relation reveals competing difference-frequency channels for every target mode, while the second gives additional sum-frequency pairs when $n\geq2$. Such multimode processes occur in SQUID-terminated resonators \cite{Wustmann:2013icg,Bengtsson:2018vvv}. The exact roots of Eq.~\eqref{eq:quantization_condition_tan} must therefore be examined together with the actual linewidths and overlap selection rules. The formulas below apply to an engineered isolated mode; otherwise, the full Hamiltonian in Eq.~\eqref{eq:multimode_rwa_hamiltonian} must be retained.

\subsection{Representative Circuit Parameters and Consistency Conditions}
\label{subsec:representative_circuit_point}

To evaluate the approximations for a representative parameter set, we take
\begin{equation}
\label{eq:circuit_scale_parameter_point}
\begin{alignedat}{2}
Z_0&=50\,{\rm \Omega},
&\quad v&=1.20\times10^8\,{\rm m\,s^{-1}},
\\
\frac{\omega_{10}}{2\pi}&=5.00\,{\rm GHz},
&\quad I_c&=1.00\,{\rm \mu A},
\\
C_J&=5.00\,{\rm fF},
&\quad \Phi_b&=0.20\Phi_0,
\\
Q_{\rm int}&=7.00\times10^3,
&\quad Q_{\rm ext}&=1.00\times10^5,
\\
T&=10\,{\rm mK}.&
\end{alignedat}
\end{equation}
The chosen resonance frequency falls within the range commonly used for superconducting coplanar resonators. We take $Q_{\rm int}=7.00\times10^3$, consistent with the single-photon value reported for a millikelvin nanoSQUID-terminated quarter-wave resonator, while the impedance and phase velocity are representative circuit values and the junction parameters are illustrative \cite{10.1063/1.3010859,Potter:2024dby}. Equations~\eqref{eq:phase_velocity}, \eqref{eq:josephson_inductance}, \eqref{eq:lsq_lj_relation}, and \eqref{eq:quantization_condition_tan} give
\begin{equation}
\label{eq:circuit_scale_derived_parameters}
\begin{alignedat}{2}
C_\ell&=1.667\times10^{-10}\,{\rm F/m},
&\; L_\ell&=4.167\times10^{-7}\,{\rm H/m},
\\
L_{\rm SQ}^{(0)}&=0.488\,{\rm mm},
&\; L_0&=5.514\,{\rm mm},
\\
C_c&=1.78\,{\rm fF},
&\; \frac{\omega_p}{2\pi}&=157.8\,{\rm GHz},
\\
Q&=6.54\times10^3,
&\; \frac{\kappa_1}{2\pi}&=0.764\,{\rm MHz},
\\
\bar n_{\rm th}&=3.79\times10^{-11},
&\; k_1L_{\rm SQ}^{(0)}&=0.128,
\end{alignedat}
\end{equation}
where $C_c$ follows from the weak-coupling estimate for the exact fundamental mode,
\begin{equation}
\label{eq:external_Q_weak_coupling}
Q_{\rm ext}
\simeq
\frac{
C_\ell L_0
\left[
\frac12+
\frac{\sin(2k_1L_0)}{4k_1L_0}
\right]
}
{\omega_{10}C_c^2Z_0}.
\end{equation}
The value of $\bar n_{\rm th}$ is obtained from Eq.~\eqref{eq:thermal_photon_number_detection}, and at $\Omega_{\rm GW}=2\omega_{10}$ the chosen parameters satisfy $\omega_{10}C_cZ_0=2.80\times10^{-3}$, $\omega_{10}/\omega_p=0.0317$, and $\Omega_{\rm GW}/\omega_p=0.0634$. For the same parameter set, the second exact root is $\omega_2/(2\pi)=15.048930\,{\rm GHz}$. Assuming the same loaded $Q$ for the first two modes, the separation from the nearest difference-frequency channel is characterized by
\begin{align}
\label{eq:circuit_scale_multimode_detuning}
\frac{\left|(\omega_2-\omega_1)-2\omega_1\right|}{2\pi}
&=48.930\,{\rm MHz},
\\
\frac{\max(\kappa_1,\kappa_2)}{2\pi}
&=2.300\,{\rm MHz},
\\
\frac{\left|(\omega_2-\omega_1)-2\omega_1\right|}
{\max(\kappa_1,\kappa_2)}
&=21.3.
\end{align}
For the fundamental mode, no additional sum-frequency pair occurs at $2\omega_1$, while the nearest difference-frequency channel remains spectrally separated for this parameter set provided that $|g_{12}|\ll |(\omega_2-\omega_1)-2\omega_1|$. In the maximally projected long-wavelength reference of Eq.~\eqref{eq:ideal_pair_and_frequency_kernels}, we obtain $P_{L,1}=\eta_{1,{\rm ref}}^{\rm pair}=0.9199$.

\subsection{Mode Equation and Mathieu Dynamics}
\label{sec:mode_equation_mathieu}
Within the restricted frequency-only approximation, where $\mathcal C_n\simeq0$, the phase of $R_n^{\rm pair}$ can be absorbed into the choice of time origin, allowing the modulation to be written as
\begin{equation}
\label{eq:pair_modulation_definition}
h_n^{\rm pair}(t)
=
\eta_n^{\rm pair}h_0
\cos(\Omega_{\rm GW}t).
\end{equation}
The equation of motion for the isolated mode then becomes
\begin{equation}
\label{eq:single_mode_eom}
\ddot q_n
+
\omega_{n0}^2
\left[1-h_n^{\rm pair}(t)\right]q_n=0.
\end{equation}
This has the form of a Mathieu equation, with its principal resonance occurring near
\begin{equation}
\label{eq:gw_parametric_resonance_condition}
    \Omega_{\rm GW}\simeq2\omega_{n0}.
\end{equation}
Introducing the detuning $\Delta_n=\Omega_{\rm GW}/2-\omega_{n0}$ and applying the rotating-wave approximation, the lossless slow-amplitude eigenvalues are
\begin{equation}
\label{eq:lossless_mathieu_eigenvalues}
\mu_\pm
=
\pm\sqrt{
\left(\frac{\eta_n^{\rm pair}h_0\omega_{n0}}{4}\right)^2
-
\Delta_n^2
}.
\end{equation}
The corresponding lossless instability band is $|\Delta_n|<\eta_n^{\rm pair}h_0\omega_{n0}/4$. If $\mathcal C_n$ or the off-diagonal kernels cannot be neglected, the appropriate starting point is the full modal Hamiltonian in Eq.~\eqref{eq:general_modal_hamiltonian}, rather than a single Mathieu equation inferred from the eigenfrequency alone.

\section{Quantum Hamiltonian and Photon-Pair Generation}
\label{sec:quantum_hamiltonian_photon_pair}

We now quantize an isolated cavity mode satisfying the spectral conditions in Eq.~\eqref{eq:single_mode_isolation_condition}. The unperturbed Hamiltonian is
\begin{equation}
\label{eq:unperturbed_hamiltonian_quantum_section}
H_{0,n}
=
\hbar\omega_{n0}
\left(a_n^\dagger a_n+\frac12\right),
\qquad
[a_n,a_n^\dagger]=1.
\end{equation}
From Eqs.~\eqref{eq:general_modal_hamiltonian} and
\eqref{eq:Rpair_definition}, and within the rotating-wave approximation
(RWA), in which rapidly oscillating nonresonant terms are neglected,
the resonant diagonal pair-interaction Hamiltonian in the interaction
picture is
\begin{equation}
\label{eq:rwa_hamiltonian_quantum_section}
\begin{split}
H_{\rm int}^{\rm RWA}(t)
={}&
\frac{\hbar\omega_{n0}h_0}{4}
\Bigl[
R_n^{\rm pair}a_n^{\dagger2}e^{-i\Delta_{\rm GW}t}
\\
&\hspace{2.5cm}+
R_n^{{\rm pair}*}a_n^2e^{i\Delta_{\rm GW}t}
\Bigr],
\end{split}
\end{equation}
where,
\begin{equation}
\label{eq:detuning_quantum_section}
\Delta_{\rm GW}
=
\Omega_{\rm GW}-2\omega_{n0}
\end{equation}
is the detuning from the degenerate parametric resonance.
Equivalently, one can write
\begin{equation}
\label{eq:dpa_hamiltonian_quantum_section}
H_{\rm int}^{\rm RWA}(t)
=
 i\hbar
\left[
\lambda_n a_n^{\dagger2}e^{-i\Delta_{\rm GW}t}
-
\lambda_n^*a_n^2e^{i\Delta_{\rm GW}t}
\right],
\end{equation}
with
\begin{equation}
\label{eq:lambda_quantum_section}
\lambda_n
=
-\frac{i\omega_{n0}h_0}{4}R_n^{\rm pair},
\qquad
|\lambda_n|
=
\frac{\eta_n^{\rm pair}h_0\omega_{n0}}{8}.
\end{equation}
All geometric and SQUID-participation factors are already contained in $R_n^{\rm pair}$, and no second dilution factor appears in Eq.~\eqref{eq:lambda_quantum_section}.

Within the frequency-only approximation, the parametric modulation experienced by mode $n$ can be expressed as
\begin{equation}
\label{eq:metric_perturb}
h_n^{\rm pair}(t)
=
2\epsilon_n^{\rm pair}
\cos(\Omega_{\rm GW}t),
\qquad
\epsilon_n^{\rm pair}
=
\frac{\eta_n^{\rm pair}h_0}{2}.
\end{equation}
With this form of the modulation, the interaction strength given in Eq.~\eqref{eq:lambda_quantum_section} is recovered directly from the oscillator Hamiltonian
$p_n^2/2+\omega_{n0}^2[1-h_n^{\rm pair}(t)]q_n^2/2$.

At exact resonance, the corresponding Heisenberg equations take the simple form
\begin{equation}
\label{eq:lossless_heisenberg_equations}
\dot a_n=2\lambda_na_n^\dagger,
\qquad
\dot a_n^\dagger=2\lambda_n^*a_n.
\end{equation}
If the cavity is initially in the vacuum state and losses are neglected, these equations lead to
\begin{equation}
\label{eq:photon_number_quantum_section}
N_n(t)
=
\sinh^2\!\left(2|\lambda_n|t\right).
\end{equation}
At sufficiently short times, such that $2|\lambda_n|t\ll1$, this expression reduces to
\begin{equation}
\label{eq:photon_number_short_time_quantum_section}
N_n(t)
\simeq
4|\lambda_n|^2t^2.
\end{equation}
This result should be understood conditionally, since the gravitational-wave strain enters through the pair-response kernel $R_n^{\rm pair}$, whose value cannot be determined from the one-dimensional circuit model alone.

Finally, under the degenerate parametric resonance condition, the generated photon pairs are centered around the cavity-mode frequency,
\begin{equation}
\label{eq:generated_photon_frequency_quantum_section}
\omega_\gamma
=\omega_{n0}
\simeq
\frac{\Omega_{\rm GW}}{2}.
\end{equation}

\subsection{Dissipation, Quality Factor, and Conditional Scaling}
\label{sec:dissipation_quality_sensitivity}

The loaded quality factor defines the total energy-decay rate
\begin{equation}
\label{eq:cavity_linewidth_quality_factor}
\kappa
=
\kappa_{\rm int}+\kappa_{\rm ext}
=
\frac{\omega_{n0}}{Q}.
\end{equation}
where $Q$ is the loaded quality factor, including both the internal cavity losses and the external coupling to the readout line. To incorporate dissipation and the coupling of the cavity mode to its environment, we work in a frame rotating at half the gravitational-wave frequency and introduce the detuning
\begin{equation}
\label{eq:rotating_frame_detuning}
\delta_n
=\omega_{n0}-\frac{\Omega_{\rm GW}}{2}
=-\frac{\Delta_{\rm GW}}{2}.
\end{equation}
In this rotating frame, the dynamics of mode $n$ are governed by the quantum Langevin equation
\begin{equation}
\label{eq:damped_langevin_equation}
\dot a_n
=
-\left(\frac{\kappa}{2}+i\delta_n\right)a_n
+2\lambda_na_n^\dagger
+\sqrt{\kappa_{\rm ext}}\,a_{{\rm in},n}
+\sqrt{\kappa_{\rm int}}\,b_{{\rm in},n},
\end{equation}
where, the last two terms describe the noise entering through the external readout channel and the internal loss channel, respectively. The eigenvalues associated with the homogeneous part of this equation are
\begin{equation}
\label{eq:damped_drift_eigenvalues}
s_\pm
=-\frac{\kappa}{2}
\pm
\sqrt{4|\lambda_n|^2-\delta_n^2}.
\end{equation}
Requiring the real parts of both eigenvalues to remain negative leads to the linear stability condition
\begin{equation}
\label{eq:general_stability_condition}
4|\lambda_n|^2
<
\delta_n^2+\frac{\kappa^2}{4}.
\end{equation}
At exact resonance, where $\delta_n=0$, the competition between parametric amplification and cavity damping can be summarized by introducing the dimensionless gain parameter
\begin{equation}
\label{eq:dimensionless_gw_gain}
\mathcal G_n
\equiv
\frac{4|\lambda_n|}{\kappa}
=
\frac{\eta_n^{\rm pair}h_0Q}{2}.
\end{equation}
The system remains below the resonant instability threshold for $\mathcal G_n<1$, while $\mathcal G_n=1$ marks the threshold itself.

The external and internal noise channels can be combined into an effective bath occupation defined by
\begin{equation}
\label{eq:effective_bath_occupation}
\bar n_{\rm b}
=
\frac{\kappa_{\rm ext}\bar n_{\rm ext}
+\kappa_{\rm int}\bar n_{\rm int}}
{\kappa}.
\end{equation}
After choosing the phase convention such that $\lambda_n$ is real and positive, the exact resonant steady-state occupation below threshold, derived in Appendix~\ref{app:gaussian_dynamics}, is
\begin{equation}
\label{eq:max_photon_number_scaling}
N_n^{\rm ss}
=
\frac{\bar n_{\rm b}+\mathcal G_n^2/2}
{1-\mathcal G_n^2}.
\end{equation}
When both input channels are in the vacuum state, this expression reduces to
\begin{equation}
\label{eq:vacuum_ss_photon_number}
N_{n,0}^{\rm ss}
=
\frac{\mathcal G_n^2}
{2(1-\mathcal G_n^2)}.
\end{equation}
In the weak-gain regime, $\mathcal G_n\ll1$, the leading contribution is therefore
\begin{equation}
\label{eq:weak_gain_photon_number}
N_{n,0}^{\rm ss}
\simeq
\frac{\mathcal G_n^2}{2}
=
\frac{(\eta_n^{\rm pair})^2h_0^2Q^2}{8},
\end{equation}
also, the same steady-state solution gives the intracavity variances of the amplified and squeezed quadratures as
\begin{equation}
\label{eq:ss_quadrature_variances}
V_{\rm amp}^{\rm ss}
=
\frac{\bar n_{\rm b}+1/2}{1-\mathcal G_n},
\qquad
V_{\rm sq}^{\rm ss}
=
\frac{\bar n_{\rm b}+1/2}{1+\mathcal G_n}.
\end{equation}

For the representative circuit parameters introduced in Eq.~\eqref{eq:circuit_scale_parameter_point}, taking the ideal-reference response $\eta_{1,{\rm ref}}^{\rm pair}=0.9199$ together with a gravitational-wave strain $h_0=10^{-21}$ gives
\begin{equation}
\label{eq:numerical_photon_number_estimate}
\begin{aligned}
\mathcal G_1=3.01\times10^{-18},\quad
N_{1,0}^{\rm ss}=4.53\times10^{-36}.
\end{aligned}
\end{equation}
The excess intracavity population leaves the cavity through the external port, producing an emitted photon flux
\begin{equation}
\label{eq:output_photon_flux}
\Phi_{\rm out}^{\rm ex}
=
\kappa_{\rm ext}N_{n,0}^{\rm ss}.
\end{equation}
For the same representative point, Eq.~\eqref{eq:output_photon_flux} yields
$\Phi_{\rm out}^{\rm ex}=1.42\times10^{-30}\,{\rm s^{-1}}$. The corresponding ideal-reference signal occupation is thus far below the equilibrium thermal occupation reported in Eq.~\eqref{eq:circuit_scale_derived_parameters}. Converting this conditional intracavity estimate into a detector-level strain sensitivity would additionally require the collection efficiency, measurement bandwidth, integration time, thermal backgrounds, amplifier-added noise, and relevant technical-noise spectra.

At exact resonance, the threshold condition may also be expressed in terms of the quality factor required to overcome cavity damping:
\begin{equation}
\label{eq:quality_factor_threshold}
Q_{\rm th}
=
\frac{2}{\eta_n^{\rm pair}h_0}.
\end{equation}
For $h_0=10^{-21}$ and $\eta_{1,{\rm ref}}^{\rm pair}=0.9199$, this gives $Q_{\rm th}=2.17\times10^{21}$. This value should be regarded as a conditional benchmark for the underlying parametric dynamics rather than as a detector-sensitivity prediction, since a quantitative assessment still requires a device-specific determination of the pair-response kernel and a complete treatment of the detector noise budget.

\section{Detection Channels}
\label{sec:detection_channels}

The response kernels categorize observables into three classes. Spontaneous pair production and squeezing are governed by \( R_n^{\text{pair}} \). On the other hand, dispersive frequency sidebands are influenced by \( R_n^{\omega} \). A seeded degenerate measurement relies on \( R_n^{\text{pair}} \), while flux-pumped down-conversion is linked to the flux derivative of \( \omega_n R_n^{\omega} \). By keeping these quantities distinct, we ensure that a frequency calibration is not confused with a comprehensive prediction of vacuum pair creation.

\subsection{Spontaneous DCE-Like Photon-Pair Production}
\label{subsec:spontaneous_dce_photon_pair}

The direct quantum signature consists of a correlated pair population near \(\Omega_{\rm GW} = 2\omega_{n0}\). In the isolated mode and weak gain limit, the vacuum contribution is described by Eq.~\eqref{eq:weak_gain_photon_number}. The thermal occupation is given by the formula:
\begin{equation}
\label{eq:thermal_photon_number_detection}
\bar n_{\rm th}
=
\frac{1}
{\exp(\hbar\omega_{n0}/k_{\rm B}T)-1}.
\end{equation}
For \(\omega_{n0}/2\pi = 5\,{\rm GHz}\) and \(T = 10\,{\rm mK}\), we find that \(\hbar \omega_{n0} / k_{\rm B} T \simeq 24\), resulting in \(\bar{n}_{\rm th} \simeq 4 \times 10^{-11}\). However, nonequilibrium photons and readout noise may still dominate over the conditional gravitational wave (GW) signal. The background-subtracted output flux is expressed as:
\begin{equation}
\label{eq:spontaneous_output_flux_detection}
\Phi_{\rm out}^{\rm ex}
=
\kappa_{\rm ext}N_{n,0}^{\rm ss}.
\end{equation}
The quadratic relationship between this output flux and \(h_0\) makes direct counting challenging for conventional strains unless the pair-response kernel is significantly enhanced and accurately calibrated.

\subsection{Photon-Pair Correlations and Squeezing}
\label{subsec:photon_pair_correlations_squeezing}

For a monochromatic parametric modulation with angular frequency
$\Omega_{\rm GW}$, we consider a state that is stationary in the frame
rotating at $\Omega_{\rm GW}/2$. Adopting the Fourier convention
\begin{equation}
a_{\rm out}(t)
=
\int_{-\infty}^{\infty}
\frac{d\omega}{2\pi}\,
e^{-i\omega t}a_{\rm out}(\omega),
\end{equation}
one obtains the following anomalous two-frequency correlation for the
outgoing field, as derived in
Appendix~\ref{app:anomalous_output_correlations}:
\begin{equation}
\label{eq:anomalous_output_correlator_detection}
\left\langle
a_{\rm out}(\omega)
a_{\rm out}(\omega')
\right\rangle
=
2\pi
\delta\!\left(
\omega+\omega'-\Omega_{\rm GW}
\right)
M_{\rm out}(\omega).
\end{equation}
Here, $a_{\rm out}(\omega)$ and $a_{\rm out}(\omega')$ describe the two
correlated components of the outgoing field. The Dirac delta function
enforces the pair-frequency condition
$\omega+\omega'=\Omega_{\rm GW}$, while the complex function
$M_{\rm out}(\omega)$ carries both the amplitude and phase of the
corresponding correlation. It is therefore natural to define the pair
spectrum as the coefficient of the Dirac distribution in
Eq.~\eqref{eq:anomalous_output_correlator_detection}:
\begin{equation}
\label{eq:pair_spectrum_detection}
C_{\rm pair}(\omega)
\equiv
M_{\rm out}(\omega).
\end{equation}

Since Eq.~\eqref{eq:anomalous_output_correlator_detection} is a
distributional relation, simply setting
$\omega'=\Omega_{\rm GW}-\omega$ does not yield a finite observable.
A physically well-defined quantity can instead be constructed by
introducing the normalized filtered output modes
\begin{align}
\label{eq:filtered_output_modes_detection}
A_f
&=
\int_{-\infty}^{\infty}
\frac{d\omega}{2\pi}
 f(\omega)a_{\rm out}(\omega),
&
\int_{-\infty}^{\infty}
\frac{d\omega}{2\pi}|f(\omega)|^2
&=1,
\\
A_g
&=
\int_{-\infty}^{\infty}
\frac{d\omega}{2\pi}
 g(\omega)a_{\rm out}(\omega),
&
\int_{-\infty}^{\infty}
\frac{d\omega}{2\pi}|g(\omega)|^2
&=1.
\end{align}
Their anomalous correlation is then finite and takes the form
\begin{equation}
\label{eq:finite_pair_correlation_detection}
\left\langle A_fA_g\right\rangle
=
\int_{-\infty}^{\infty}
\frac{d\omega}{2\pi}
 f(\omega)g(\Omega_{\rm GW}-\omega)
M_{\rm out}(\omega).
\end{equation}
Thus, $C_{\rm pair}(\omega)$ characterizes the phase-sensitive
correlation between the conjugate output frequencies $\omega$ and
$\Omega_{\rm GW}-\omega$. Its explicit form is obtained from the
input--output scattering matrix in
Appendix~\ref{app:anomalous_output_correlations}. In the weak-gain
limit, Eqs.~\eqref{eq:appB_explicit_output_spectrum} and
\eqref{eq:appB_explicit_anomalous_spectrum} give
$S_{\rm out}^{(0)}=\mathcal O(h_0^2)$ and
$M_{\rm out}=\mathcal O(h_0)$. A nonzero pair spectrum therefore
provides a phase-sensitive signature of coherent pair generation,
rather than merely indicating an incoherent excess photon population.

For a normalized temporal or spectral output mode, the corresponding
quadrature operator may be written as
\begin{equation}
\label{eq:output_quadrature_detection}
X_\theta
=
\frac{1}{\sqrt2}
\left(a_{\rm out}e^{-i\theta}
+a_{\rm out}^\dagger e^{i\theta}\right).
\end{equation}
Taking the vacuum variance to be $1/2$, quadrature squeezing occurs
whenever
\begin{equation}
\label{eq:quadrature_squeezing_condition_detection}
\left\langle\Delta X_\theta^2\right\rangle
<\frac12.
\end{equation}
In practice, however, internal loss and finite detection efficiency
mix unsqueezed environmental modes into the measured output and
thereby reduce the observable squeezing, as discussed in
Appendix~\ref{app:output_quadrature_squeezing}.

\subsection{Coherent-Probe Transduction}
\label{subsec:sideband_detection_coherent_drive}

A coherent probe provides a direct way to convert the resonance-frequency modulation into a field response that is linear in $h_0$. Denoting the drive frequency by $\omega_d$ and defining $\Delta_d=\omega_d-\omega_{n0}$, with $\theta_n^\omega=\arg R_n^\omega$, Eq.~\eqref{eq:frequency_response_definition} gives
\begin{equation}
\label{eq:gw_frequency_modulation_detection}
\delta\omega_n^{\rm GW}(t)
=
\delta\omega_{{\rm GW},n}
\cos(\Omega_{\rm GW}t-\theta_n^\omega).
\end{equation}
The modulation amplitude follows from Eq.~\eqref{eq:gw_frequency_shift_amplitude} as
\begin{equation}
\label{eq:gw_frequency_shift_amplitude_detection}
\delta\omega_{{\rm GW},n}
=
\frac{\eta_n^\omega h_0\omega_{n0}}{2}.
\end{equation}
For a sideband displaced from the drive by an angular frequency $\nu$, the cavity susceptibility is
\begin{equation}
\label{eq:cavity_susceptibility_detection}
\chi_c(\nu)
=
\frac{1}{\kappa/2-i(\Delta_d+\nu)}.
\end{equation}
Keeping only terms that are first order in the modulation, the relative sideband amplitudes are therefore
\begin{equation}
\label{eq:sideband_amplitude_ratio_general_detection}
\frac{|\delta\alpha_\pm|}{|\alpha_0|}
=
\frac{\delta\omega_{{\rm GW},n}}{2}
|\chi_c(\pm\Omega_{\rm GW})|.
\end{equation}
For resonant probing in the regime $\Omega_{\rm GW}\gg\kappa$, this expression reduces to
\begin{equation}
\label{eq:sideband_amplitude_ratio_strain_detection}
\frac{|\delta\alpha_\pm|}{|\alpha_0|}
\simeq
\frac{\eta_n^\omega h_0\omega_{n0}}
{4\Omega_{\rm GW}}.
\end{equation}
At $\Omega_{\rm GW}\simeq2\omega_{n0}$, however, these components should not be interpreted as an ordinary pair of resonantly enhanced sidebands of a single cavity mode. Rather, the expression describes a far-detuned upper component near $3\omega_{n0}$ together with its conjugate negative-frequency component.

Near the degenerate parametric condition, the role of the coherent probe changes, as it now acts as a seed for the pair interaction. The corresponding mean field satisfies
\begin{equation}
\label{eq:seeded_parametric_response_detection}
\dot\alpha
=
-\left(\frac{\kappa}{2}+i\delta_n\right)\alpha
+2\lambda_n\alpha^*
+\sqrt{\kappa_{\rm ext}}\,\alpha_{\rm in}.
\end{equation}
At exact resonance, the phase may be chosen such that $\lambda_n>0$, in which case the two mean quadratures obey
\begin{align}
\label{eq:seeded_quadrature_equations}
\dot X
&=-\frac{\kappa}{2}(1-\mathcal G_n)X
+\sqrt{\kappa_{\rm ext}}X_{\rm in},
\\
\dot P
&=-\frac{\kappa}{2}(1+\mathcal G_n)P
+\sqrt{\kappa_{\rm ext}}P_{\rm in}.
\end{align}
Their steady-state responses, normalized to the corresponding unpumped intracavity values, are then
\begin{equation}
\label{eq:seeded_quadrature_gain}
\frac{X_{\rm ss}}{X_{\rm ss}^{(0)}}
=
\frac{1}{1-\mathcal G_n},
\qquad
\frac{P_{\rm ss}}{P_{\rm ss}^{(0)}}
=
\frac{1}{1+\mathcal G_n}.
\end{equation}
Thus, in the weak-gain regime $\mathcal G_n\ll1$, the two quadratures acquire fractional changes of $\pm\mathcal G_n$, both of which are linear in $h_0$.

\subsection{Flux-Pumped Frequency Conversion and the Resonant Pump Background}
\label{subsec:squid_heterodyne_transduction}
Rather than assuming a SQUID flux generated directly by the gravitational wave, we use the external pump to modulate the flux dependence of the detector response introduced above. The applied flux is taken to be
\begin{equation}
\label{eq:flux_pump_definition}
\Phi(t)
=
\Phi_b
+
\Phi_p\cos(\Omega_pt+\varphi_p).
\end{equation}
Away from the half-flux bias point, Eqs.~\eqref{eq:lsq_flux} and \eqref{eq:exact_frequency_variation} give
\begin{align}
\label{eq:flux_derivative_LSQ_omega}
\frac{\partial\ln L_{\rm SQ}}{\partial\Phi}
&=
\frac{\pi}{\Phi_0}
\tan\left(\frac{\pi\Phi}{\Phi_0}\right),
\\
\frac{\partial\omega_n}{\partial\Phi}
&=
-\omega_nP_{{\rm SQ},n}
\frac{\pi}{\Phi_0}
\tan\left(\frac{\pi\Phi}{\Phi_0}\right).
\end{align}
Consequently, even in the absence of a gravitational wave, the pump induces the frequency modulation
\begin{equation}
\label{eq:pump_only_frequency_modulation}
\delta\omega_n^{(p)}(t)
=
\Phi_p
\left.
\frac{\partial\omega_n}{\partial\Phi}
\right|_{\Phi_b}
\cos(\Omega_pt+\varphi_p).
\end{equation}

To describe the flux dependence of the gravitational-wave response, it is convenient to introduce
\begin{equation}
\label{eq:F_response_definition}
\mathcal F_n(\Phi)
\equiv
\omega_n(\Phi)R_n^\omega(\Phi).
\end{equation}
In the quasistatic-bias limit, the gravitational-wave-induced frequency perturbation evaluated at the instantaneous bias can then be written as
\begin{equation}
\label{eq:gw_frequency_response_flux_dependent}
\delta\omega_n^{\rm GW}(t;\Phi)
=
\operatorname{Re}\!\left[
\mathcal F_n(\Phi)h_0e^{-i\Omega_{\rm GW}t}
\right].
\end{equation}
Expanding this expression to first order in $\Phi_p$, one obtains the cross term
\begin{equation}
\label{eq:flux_response_cross_term}
\delta\omega_n^{\rm cross}(t)
=
\Phi_p\cos(\Omega_pt+\varphi_p)
\operatorname{Re}\!\left[
\mathcal F_n'(\Phi_b)h_0e^{-i\Omega_{\rm GW}t}
\right].
\end{equation}
This term contains both sum- and difference-frequency components, with the latter having the amplitude
\begin{equation}
\label{eq:heterodyne_frequency_shift_detection}
\delta\omega_{\rm IF}
=
\frac{\Phi_ph_0}{2}
\left|
\frac{\partial}{\partial\Phi}
\left[\omega_n(\Phi)R_n^\omega(\Phi)\right]_{\Phi_b}
\right|,
\end{equation}
at the intermediate frequency
\begin{equation}
\label{eq:intermediate_frequency_detection}
\Omega_{\rm IF}
=|\Omega_{\rm GW}-\Omega_p|.
\end{equation}
For a coherent probe, the corresponding sideband response is
\begin{equation}
\label{eq:heterodyne_sideband_amplitude_detection}
\frac{|\delta\alpha_{{\rm IF},\pm}|}{|\alpha_0|}
=
\frac{\delta\omega_{\rm IF}}{2}
|\chi_c(\pm\Omega_{\rm IF})|.
\end{equation}

When the pump frequency approaches $\Omega_p=2\omega_{n0}$, however, the pump-only modulation in Eq.~\eqref{eq:pump_only_frequency_modulation} also generates the parametric coupling
\begin{equation}
\label{eq:direct_flux_pump_pair_coupling}
|\lambda_n^{(p)}|
=
\frac{|\Phi_p|}{4}
\left|
\left.
\frac{\partial\omega_n}{\partial\Phi}
\right|_{\Phi_b}
\right|.
\end{equation}
Retaining only the derivative-based mixing contribution is therefore consistent only when the pump remains well outside the corresponding parametric linewidth,
\begin{equation}
\label{eq:off_resonant_flux_pump_condition}
|\Omega_p-2\omega_{n0}|
\gg
\max\left\{\kappa,4|\lambda_n^{(p)}|\right\},
\end{equation}
while also satisfying the multimode conditions in Eq.~\eqref{eq:single_mode_isolation_condition} after replacing $\Omega_{\rm GW}$ by $\Omega_p$. For a fixed pump amplitude, the cross term is linear in $h_0$ and vanishes unless $\partial_\Phi(\omega_nR_n^\omega)$ is nonzero. Equation~\eqref{eq:heterodyne_frequency_shift_detection} applies in the quasistatic regime, where $|\Phi_p|$ is small, the pump lies below the relevant mechanical and electromagnetic response poles, and Eq.~\eqref{eq:off_resonant_flux_pump_condition} is satisfied. Although $\Omega_p\ll\omega_p$ suppresses junction-plasma dynamics, it does not by itself make the cavity response adiabatic. In particular, choosing $\Omega_p\simeq\Omega_{\rm GW}\simeq2\omega_{n0}$ to produce a small intermediate frequency violates Eq.~\eqref{eq:off_resonant_flux_pump_condition}, in which case the pump-induced parametric background must be included through a Floquet input--output calculation \cite{Wilson:2011rsw,Wustmann:2013icg,Bengtsson:2018vvv}. Pump leakage, heating, flux noise, and ordinary intermodulation may also contribute additional backgrounds.

\subsection{Detector Response and Experimental Considerations}
\label{sec:detector_response}

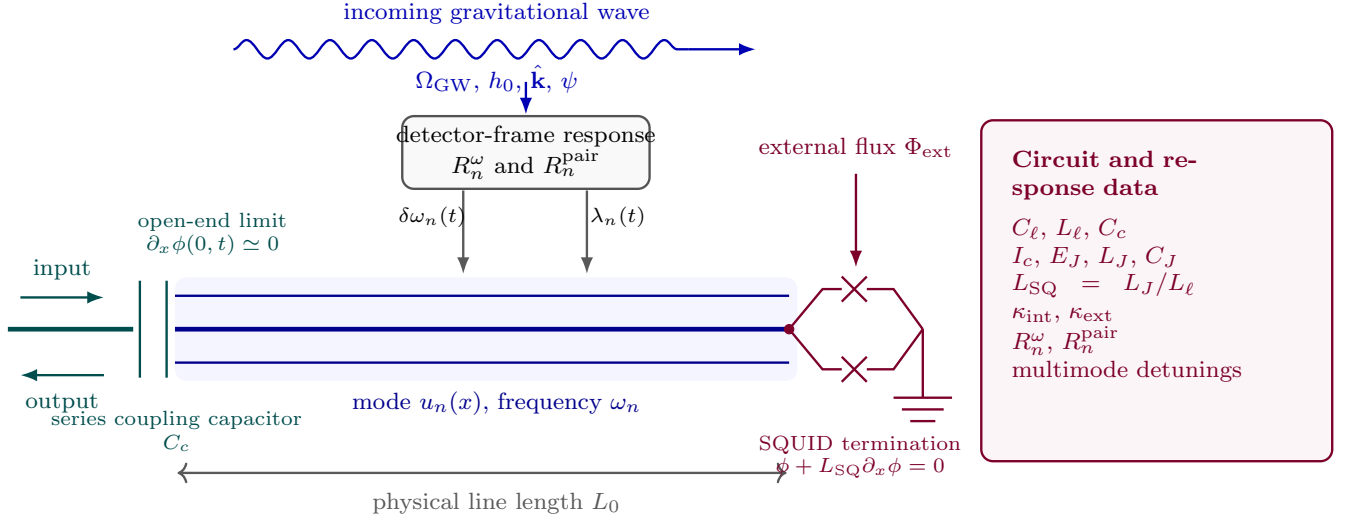
\begin{figure*}[tbp]
\centering
\resizebox{0.98\textwidth}{!}{%
\begin{tikzpicture}[
    every node/.style={font=\footnotesize},
    wave/.style={decorate,decoration={snake,amplitude=1.1mm,segment length=5mm}},
    arrow/.style={-{Latex[length=2mm]},thick},
    cavity/.style={thick,blue!55!black},
    squid/.style={thick,purple!65!black},
    readout/.style={thick,teal!60!black},
    gw/.style={wave,thick,blue!70!black}
]

% Readout line and series coupling capacitor
\draw[readout,ultra thick] (-6.5,0) -- (-5.0,0);
\draw[arrow,teal!60!black] (-6.35,0.38) -- (-5.35,0.38);
\node[teal!60!black] at (-5.85,0.72) {input};
\draw[arrow,teal!60!black] (-5.35,-0.55) -- (-6.35,-0.55);
\node[teal!60!black] at (-5.85,-0.88) {output};
\draw[readout] (-4.92,0.58) -- (-4.92,-0.58);
\draw[readout] (-4.60,0.58) -- (-4.60,-0.58);
\node[teal!60!black,align=center,font=\scriptsize]
at (-4.48,-1.22) {series coupling capacitor\\$C_c$};

% Quarter-wave CPW cavity
\fill[blue!4,rounded corners] (-4.50,0.62) rectangle (2.95,-0.62);
\draw[cavity] (-4.50,0.40) -- (2.85,0.40);
\draw[cavity] (-4.50,-0.40) -- (2.85,-0.40);
\draw[cavity,ultra thick] (-4.50,0) -- (2.85,0);
\node[blue!55!black] at (-0.65,-0.88)
{mode $u_n(x)$, frequency $\omega_n$};
\node[teal!60!black,align=center,font=\scriptsize]
at (-4.05,1.14) {open-end limit\\$\partial_x\phi(0,t)\simeq0$};
\draw[<->,thick,gray!70!black] (-4.47,-1.72) -- (2.82,-1.72);
\node[gray!70!black] at (-0.65,-2.08) {physical line length $L_0$};

% SQUID loop from cavity node to ground
\filldraw[purple!65!black] (2.85,0) circle (0.055);
\draw[squid] (2.85,0) -- (3.25,0.48) -- (3.48,0.48);
\draw[squid] (3.48,0.33) -- (3.78,0.63);
\draw[squid] (3.78,0.33) -- (3.48,0.63);
\draw[squid] (3.78,0.48) -- (4.10,0.48) -- (4.45,0);
\draw[squid] (2.85,0) -- (3.25,-0.48) -- (3.48,-0.48);
\draw[squid] (3.48,-0.63) -- (3.78,-0.33);
\draw[squid] (3.78,-0.63) -- (3.48,-0.33);
\draw[squid] (3.78,-0.48) -- (4.10,-0.48) -- (4.45,0);
\draw[squid] (4.45,0) -- (4.45,-0.82);
\draw[squid] (4.10,-0.82) -- (4.80,-0.82);
\draw[squid] (4.20,-0.96) -- (4.70,-0.96);
\draw[squid] (4.30,-1.10) -- (4.60,-1.10);
\node[purple!65!black,align=center,font=\scriptsize]
at (3.70,-1.38) {SQUID termination};
\node[purple!65!black,align=center,font=\scriptsize]
at (3.70,-1.66) {$\phi+L_{\rm SQ}\partial_x\phi=0$};
\draw[arrow,purple!65!black] (3.65,1.86) -- (3.65,0.74);
\node[purple!65!black] at (3.65,2.18) {external flux $\Phi_{\rm ext}$};

% GW and response box
\draw[gw] (-3.8,3.35) -- (1.65,3.35);
\draw[arrow,blue!70!black] (1.65,3.35) -- (2.45,3.35);
\node[blue!70!black] at (-0.65,3.78) {incoming gravitational wave};
\node[blue!70!black] at (-0.65,3.00)
{$\Omega_{\rm GW},\,h_0,\,\hat{\mathbf k},\,\psi$};
\filldraw[fill=gray!5,draw=gray!65!black,rounded corners,thick]
(-1.78,1.68) rectangle (1.18,2.55);
\node[align=center] at (-0.30,2.12)
{detector-frame response\\$R_n^{\omega}$ and $R_n^{\rm pair}$};
\draw[arrow,blue!70!black] (-0.30,2.96) -- (-0.30,2.57);
\draw[arrow,gray!70!black] (-1.05,1.66) -- (-1.05,0.66);
\draw[arrow,gray!70!black] (0.43,1.66) -- (0.43,0.66);
\node[align=center,font=\scriptsize] at (-1.42,1.34) {$\delta\omega_n(t)$};
\node[align=center,font=\scriptsize] at (0.80,1.34) {$\lambda_n(t)$};

% Parameter box
\filldraw[fill=purple!4,draw=purple!65!black,rounded corners,thick]
(5.15,2.50) rectangle (9.35,-1.58);
\node[align=left,text width=3.55cm,anchor=north west,purple!65!black]
at (5.40,2.25)
{\textbf{Circuit and response data}\\[4pt]
$C_\ell,\,L_\ell,\,C_c$\\
$I_c,\,E_J,\,L_J,\,C_J$\\
$L_{\rm SQ}=L_J/L_\ell$\\
$\kappa_{\rm int},\,\kappa_{\rm ext}$\\
$R_n^{\omega},\,R_n^{\rm pair}$\\
multimode detunings};

\end{tikzpicture}%
}
\caption{Schematic of the capacitively coupled, SQUID-terminated quarter-wave resonator. The coupling capacitor produces an approximately open boundary for the closed-cavity eigenproblem, whereas the SQUID supplies the tunable Robin boundary. The gravitational perturbation is mapped to the ordinary frequency response $R_n^{\omega}$ and the pair-response kernel $R_n^{\rm pair}$; the diagram does not assume a universal displacement $hL_0/2$ or a direct gravitationally generated SQUID flux.}
\label{fig:squid_gw_detector}
\end{figure*}

Figure~\ref{fig:squid_gw_detector} summarizes the circuit geometry, in which the coupling capacitor provides an approximately open electromagnetic boundary at the readout end, while the SQUID imposes an inductive Robin boundary at the opposite end. Rather than representing the gravitational wave as a prescribed displacement of the transmission line, its effect is introduced through the detector-frame response kernels $R_n^\omega$ and $R_n^{\rm pair}$, which map the perturbation onto the resonance-frequency and pair-production channels, respectively.

Since the outgoing microwave field is the measured observable, a complete sensitivity estimate must account for $\kappa_{\rm int}$, $\kappa_{\rm ext}$, the bath occupations, amplifier-added noise, collection efficiency, integration bandwidth, pump leakage, and the calibrated response kernels, while the exact mode spectrum must also satisfy Eq.~\eqref{eq:single_mode_isolation_condition}. Once these device-specific quantities have been determined, the present analysis provides the corresponding conditional quantum dynamics.

\section{Gaussian Dynamics and Numerical Consistency Checks}
\label{sec:numerical-simulation}

In the numerical analysis, we retain the same GW-conditioned coupling $\lambda_n$ used in the analytical treatment and evolve the corresponding physical Gaussian moments, without treating the external SQUID flux as a substitute for the gravitational perturbation. The dynamics are therefore described by
\begin{equation}
\label{eq:simulation-langevin}
\dot a
=
-\left(\frac{\kappa}{2}+i\delta\right)a
+2\lambda a^\dagger
+\sqrt{\kappa_{\rm ext}}a_{\rm in}
+\sqrt{\kappa_{\rm int}}b_{\rm in},
\end{equation}
where $\lambda=\lambda_n$ is given by Eq.~\eqref{eq:lambda_quantum_section}. Accordingly, the coefficient of $a^\dagger$ remains $2\lambda$, in agreement with the Hamiltonian convention adopted throughout the paper.

To characterize the Gaussian state, we introduce the photon number and anomalous moment as
\begin{equation}
\label{eq:simulation-moments}
N(t)=\langle a^\dagger a\rangle,
\qquad
M(t)=\langle a^2\rangle.
\end{equation}
For an effective bath occupation $\bar n_{\rm b}$, their exact evolution equations take the form
\begin{align}
\label{eq:simulation-moment-equations}
\dot N
&=-\kappa(N-\bar n_{\rm b})
+2\lambda M^*+2\lambda^*M,
\\
\dot M
&=-(\kappa+2i\delta)M
+4\lambda N+2\lambda.
\end{align}
The final term in the equation for $M$ is the vacuum seed required by $[a,a^\dagger]=1$, and the resulting system preserves the Gaussian uncertainty relation when evolved using a stable matrix exponential or a sufficiently converged ODE solver.

The coherent mean, on the other hand, satisfies
\begin{equation}
\label{eq:simulation-mean-equation}
\dot\alpha
=-\left(\frac{\kappa}{2}+i\delta\right)\alpha
+2\lambda\alpha^*
+\sqrt{\kappa_{\rm ext}}\alpha_{\rm in}.
\end{equation}
Thus, in the absence of both an initial coherent amplitude and an external coherent input,
\begin{equation}
\label{eq:zero_mean_vacuum}
\alpha(0)=0,
\quad
\alpha_{\rm in}=0
\quad\Longrightarrow\quad
\alpha(t)=0.
\end{equation}
Although spontaneous pair creation changes both $N$ and $M$, it does not generate a coherent field, which can grow only when a seed or a linear drive is explicitly introduced.

At exact resonance, assuming vacuum input and real $\lambda>0$, it is convenient to define $\tau=\kappa t$ and $\mathcal G=4\lambda/\kappa<1$. The exact transient quadrature variances are then
\begin{align}
\label{eq:exact_transient_variances}
V_{\rm amp}(\tau)
&=
\frac{1-\mathcal G e^{-(1-\mathcal G)\tau}}
{2(1-\mathcal G)},
\\
V_{\rm sq}(\tau)
&=
\frac{1+\mathcal G e^{-(1+\mathcal G)\tau}}
{2(1+\mathcal G)}.
\end{align}
The corresponding photon number and anomalous moment follow as
\begin{equation}
\label{eq:exact_transient_NM}
N(\tau)
=\frac{V_{\rm amp}+V_{\rm sq}-1}{2}\ge0,
\qquad
M(\tau)
=\frac{V_{\rm amp}-V_{\rm sq}}{2}.
\end{equation}
This form makes it explicit that $N(\tau)\geq0$ is preserved throughout the evolution.

\begin{figure}[tbp]
\centering
\includegraphics[width=\columnwidth]{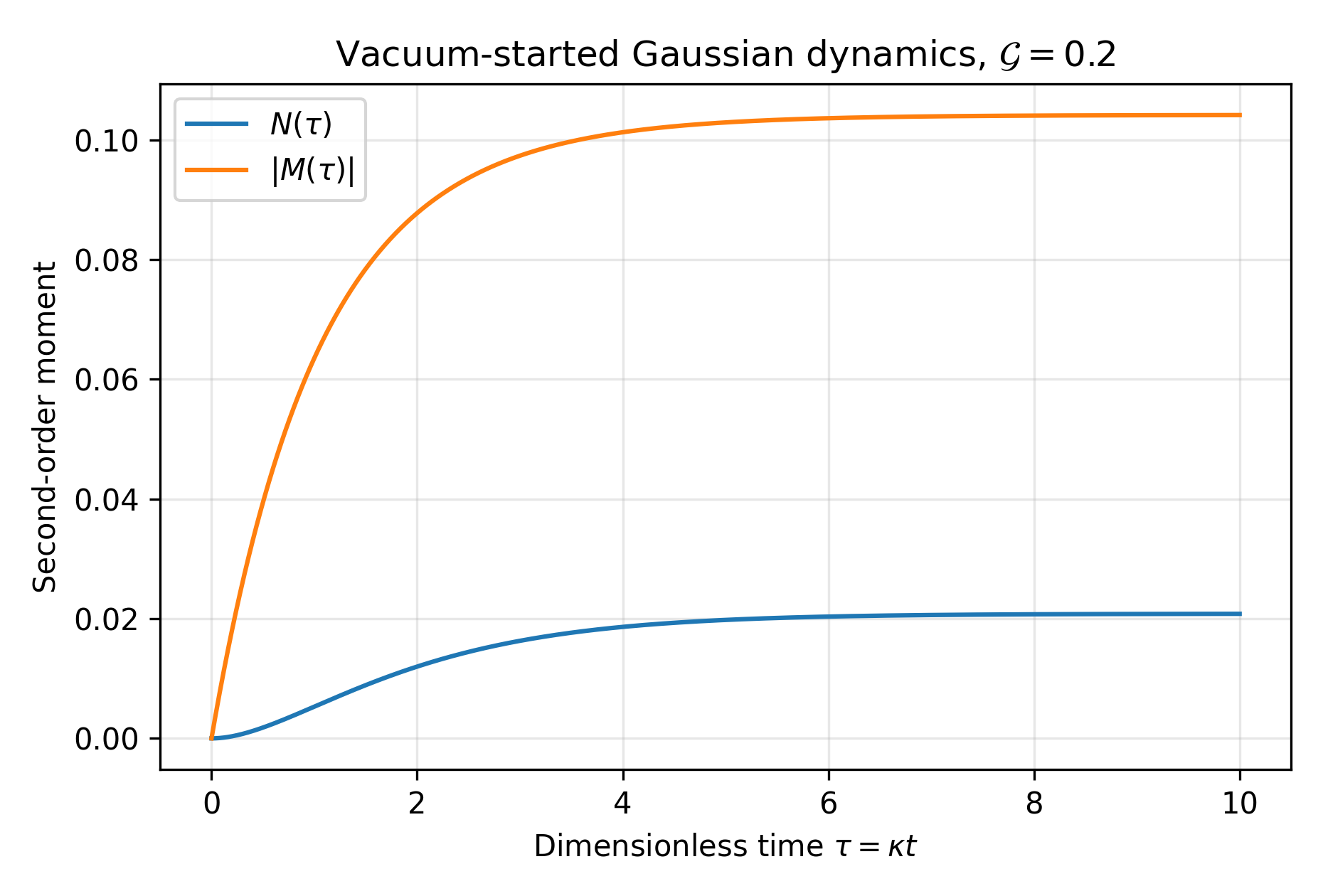}
\caption{Positivity-preserving vacuum-started evolution of the photon number $N$ and anomalous moment $|M|$ for $\mathcal G=0.2$ at exact resonance. Both moments begin at zero, and $N(\tau)$ remains nonnegative at all times.}
\label{fig:corrected-moments}
\end{figure}

\begin{figure}[t]
\centering
\includegraphics[width=\columnwidth]{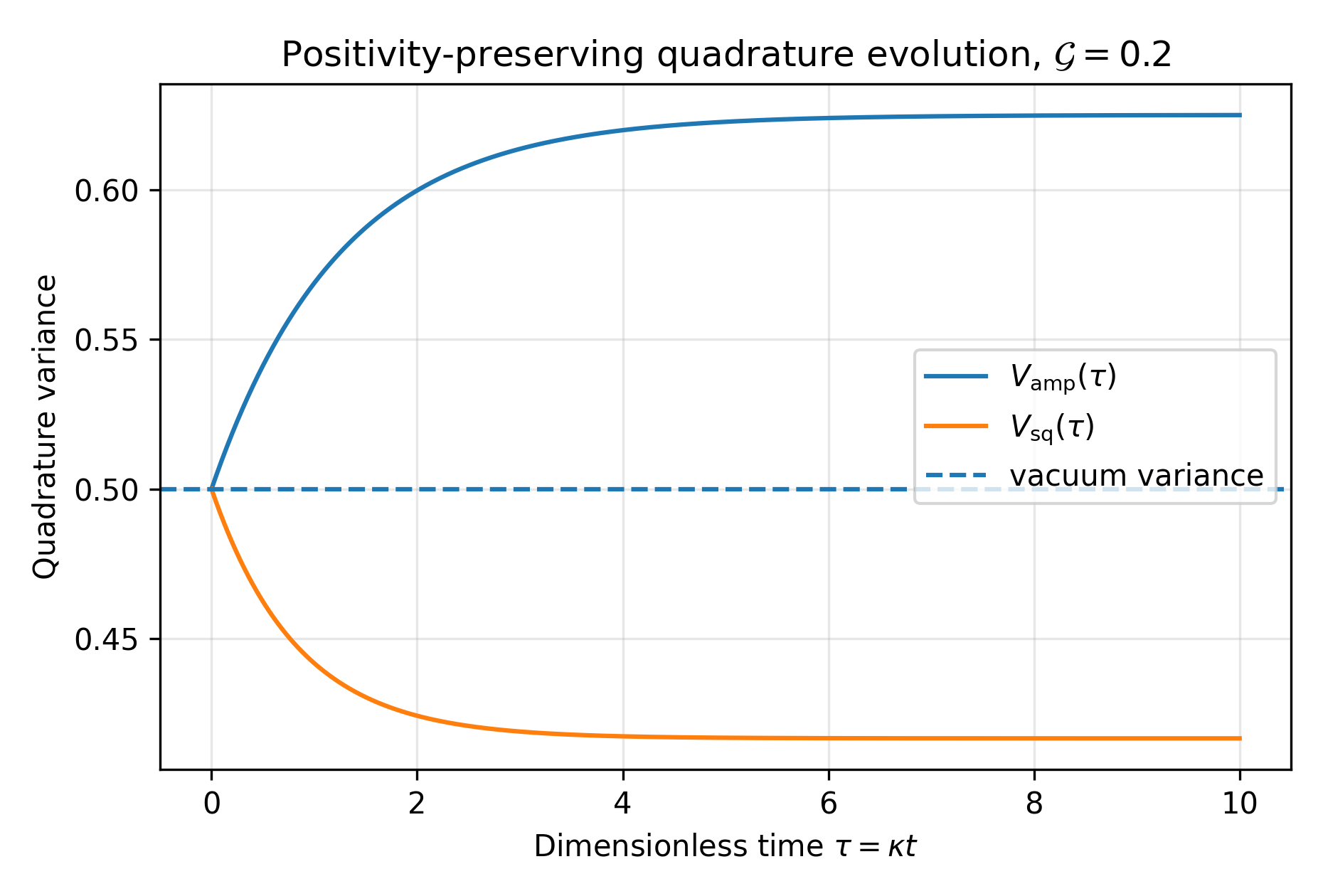}
\caption{Exact transient amplified and squeezed quadrature variances for the same benchmark. The state begins at the vacuum value $1/2$; one variance decreases and the conjugate variance increases while the uncertainty relation remains satisfied.}
\label{fig:corrected-quadratures}
\end{figure}

Figures~\ref{fig:corrected-moments} and \ref{fig:corrected-quadratures} use $\mathcal G=0.2$ solely as a visible benchmark of the linear Gaussian dynamics rather than as an astrophysical prediction. For the representative point in Eq.~\eqref{eq:circuit_scale_parameter_point} and the maximally projected ideal response, this choice would correspond to $h_0=6.65\times10^{-5}$, whereas $h_0=10^{-21}$ gives $\mathcal G=3.01\times10^{-18}$.

\begin{figure}[t]
\centering
\includegraphics[width=\columnwidth]{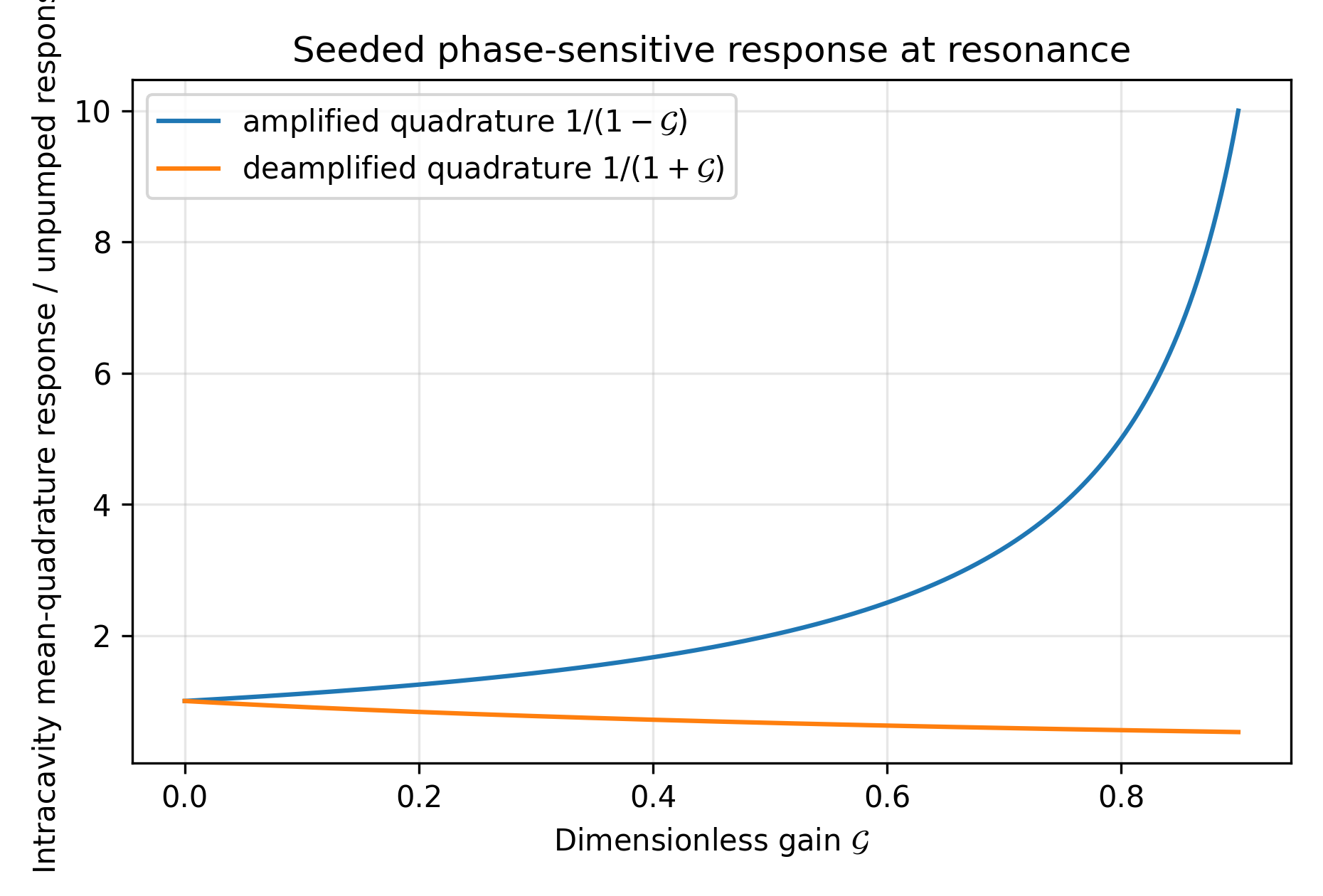}
\caption{Intracavity mean-quadrature response relative to the unpumped response for an explicitly seeded resonant mode. The amplified and deamplified responses are $1/(1-\mathcal G)$ and $1/(1+\mathcal G)$, respectively.}
\label{fig:corrected-seeded}
\end{figure}

For a coherently seeded resonant mode, Eq.~\eqref{eq:seeded_quadrature_gain} gives the phase-sensitive mean response displayed in Fig.~\ref{fig:corrected-seeded}, whereas the unseeded case is not shown because its mean field remains identically zero.

\begin{figure}[t]
\centering
\includegraphics[width=\columnwidth]{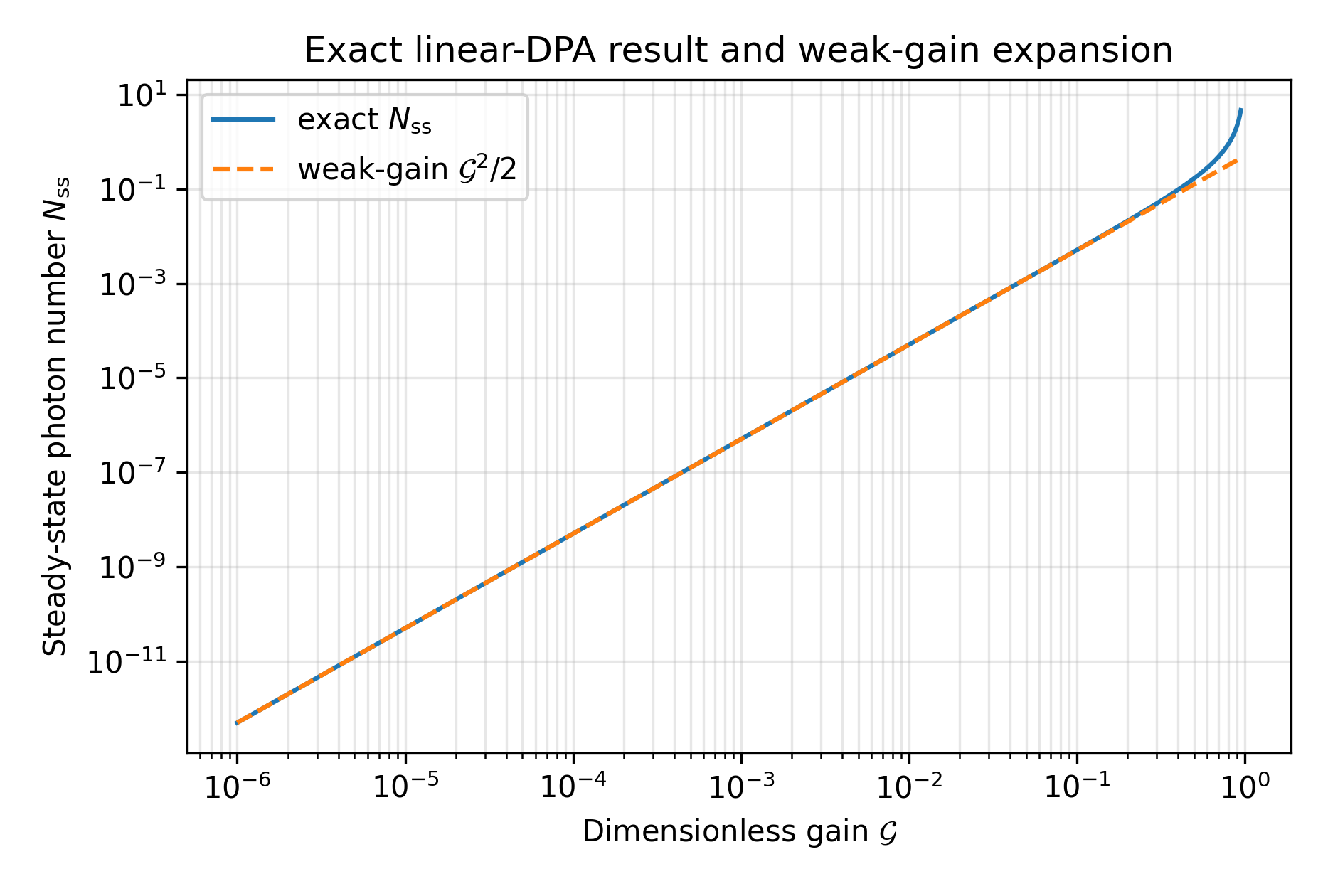}
\caption{Exact linear-DPA vacuum occupation $N_{\rm ss}=\mathcal G^2/[2(1-\mathcal G^2)]$ and the weak-gain expansion $\mathcal G^2/2$. This comparison is an analytic consistency check, not an independent numerical confirmation and not a nonlinear calculation.}
\label{fig:corrected-steady-state}
\end{figure}

Figure~\ref{fig:corrected-steady-state} compares the exact linear-DPA steady state with its weak-gain expansion, and their departure near threshold reflects the pole of the linear below-threshold solution rather than nonlinear saturation, which is not included in the present model.

Taken together, the numerical and analytical results depend on the response kernels, cavity spectrum, damping rates, bath occupations, and the specified drive or initial conditions.

\section{Discussion and Conclusion}
\label{sec:discussion_conclusion}

We have developed the conditional parametric dynamics of a capacitively coupled, SQUID-terminated superconducting cavity by combining the boundary conditions derived in Sec.~\ref{sec:squid_cavity} with an explicit separation between device-independent oscillator dynamics and device-specific gravitational transduction. Since the coupling capacitor imposes an approximately open boundary, the relevant static resonator has a quarter-wave rather than a half-wave structure, with its exact spectrum determined by $\cot(k_nL_0)=k_nL_{\rm SQ}$ and its first-order variation given by Eq.~\eqref{eq:exact_frequency_variation}.

One of the main results is the distinction between the frequency-response kernel $R_n^\omega$ and the pair-response kernel $R_n^{\rm pair}$. While the former governs the resonance-frequency sidebands, the latter determines the $a_n^{\dagger2}+a_n^2$ interaction, and the two coincide only within the restricted frequency-only approximation. A calibrated frequency shift therefore does not, in general, determine the spontaneous pair-production rate unless variations in the modal capacitance, impedance, normalization, and mode shape can be shown to be negligible.

The single-mode degenerate-parametric-amplifier formulas apply only when the spectral-isolation conditions in Eq.~\eqref{eq:single_mode_isolation_condition} are satisfied. Because the quarter-wave spectrum is nearly harmonic, possible sum- and difference-frequency channels must be examined using the exact roots, linewidths, and overlap selection rules. For the representative parameters considered in Sec.~\ref{subsec:representative_circuit_point}, the nearest difference-frequency channel is detuned by $21.3$ times the larger modal linewidth, subject to the stated bound on $g_{12}$; if this isolation is not maintained, the multimode Hamiltonian in Eq.~\eqref{eq:multimode_rwa_hamiltonian} must be retained.

For an isolated mode, the conditional pair coupling is $|\lambda_n|=\eta_n^{\rm pair}h_0\omega_{n0}/8$, implying that the vacuum steady-state occupation is quadratic in the strain, whereas the anomalous pair spectrum and an explicitly seeded phase-sensitive response are linear in the weak-gain regime. At the same time, the Gaussian dynamics preserve $N(t)\ge0$, while an unseeded vacuum state continues to satisfy $\langle a\rangle=0$. The exact steady-state curve should nevertheless be interpreted within the linear-DPA framework, since nonlinear saturation would require additional circuit nonlinearities and pump dynamics.

Flux-pumped frequency conversion can likewise be described without introducing a direct GW-induced SQUID flux. In the quasistatic regime, the derivative $\partial_\Phi[\omega_nR_n^\omega]$ determines the cross term proportional to $h_0\Phi_p$, whereas a pump near $2\omega_{n0}$ also generates the pump-induced parametric background in Eq.~\eqref{eq:direct_flux_pump_pair_coupling}. This near-resonant regime requires a Floquet input--output treatment, together with calibration of the background magnitude and phase and the inclusion of its technical contributions in the noise model.

For an ideal freely responding line in the pure-stiffness approximation, Eq.~\eqref{eq:ideal_pair_and_frequency_kernels} gives explicit nonzero values for both response kernels, while the filtered correlation in Eq.~\eqref{eq:finite_pair_correlation_detection} and the scattering solution in Appendix~\ref{app:anomalous_output_correlations} specify the corresponding measurable pair spectrum. The present work, however, does not evaluate $R_n^\omega$ or $R_n^{\rm pair}$ for a particular chip, substrate, package, or anchoring configuration and therefore does not establish an absolute gravitational-wave sensitivity. A detector-level prediction would require a proper-detector-frame elastic--electromagnetic calculation, the resulting overlap matrices $C_{mn}$ and $K_{mn}$, an experimental or simulated calibration of their flux dependence, and a complete input--output noise budget. Within this scope, the system is best regarded as a theoretical framework for a tunable narrow-band quantum transducer and for controlled laboratory studies of weak parametric modulation in superconducting circuits.

\section*{Acknowledgements}
This research is supported by the research grant of the University of Tabriz (sad/3544).

\appendix
\numberwithin{equation}{section}

\section{Exact variation of the quarter-wave eigenvalues}
\label{app:eigenvalue_variation}

The static eigenvalue condition is
\begin{equation}
\label{eq:appA_F}
F(k,L,\ell)
=
\cos(kL)-k\ell\sin(kL)=0,
\end{equation}
where $L=L_0$ and $\ell=L_{\rm SQ}$. Its derivatives are
\begin{align}
\label{eq:appA_derivatives}
F_k
&=-L\sin(kL)-\ell\sin(kL)-k\ell L\cos(kL),
\\
F_L
&=-k\sin(kL)-k^2\ell\cos(kL),
\\
F_\ell
&=-k\sin(kL).
\end{align}
On a root, $\cos(kL)=k\ell\sin(kL)$, so
\begin{align}
\label{eq:appA_derivatives_on_shell}
F_k
&=-\sin(kL)\left[L+\ell+k^2\ell^2L\right],
\\
F_L
&=-k\sin(kL)\left(1+k^2\ell^2\right).
\end{align}
Implicit differentiation, $F_k\delta k+F_L\delta L+F_\ell\delta\ell=0$, gives
\begin{equation}
\label{eq:appA_delta_k}
\frac{\delta k}{k}
=-
\frac{(1+k^2\ell^2)\delta L+\delta\ell}
{L(1+k^2\ell^2)+\ell}.
\end{equation}
Defining
\begin{equation}
\label{eq:appA_participations}
P_L
=
\frac{L(1+k^2\ell^2)}{L(1+k^2\ell^2)+\ell},
\qquad
P_\ell
=
\frac{\ell}{L(1+k^2\ell^2)+\ell},
\end{equation}
we obtain $P_L+P_\ell=1$ and
\begin{equation}
\label{eq:appA_delta_omega}
\frac{\delta\omega}{\omega}
=
\frac{\delta v}{v}
-P_L\frac{\delta L}{L}
-P_\ell\frac{\delta\ell}{\ell}.
\end{equation}
This derivation shows explicitly where the line-length and SQUID-participation factors enter. Once they have been included in $R_n^\omega$, they must not be multiplied into the response a second time.

\section{Quadratic modal projection and the two response kernels}
\label{app:quadratic_projection}

Write the quadratic modal Lagrangian in matrix notation as
\begin{equation}
\label{eq:appB_lagrangian_matrix}
\mathcal L
=
\frac12\dot{\mathbf q}^{\mathsf T}
(\mathbf I+\mathbf C)\dot{\mathbf q}
-
\frac12\mathbf q^{\mathsf T}
(\boldsymbol\Omega^2+\mathbf K)\mathbf q,
\end{equation}
where $\boldsymbol\Omega={\rm diag}(\omega_{10},\omega_{20},\ldots)$. The canonical momentum is $\mathbf p=(\mathbf I+\mathbf C)\dot{\mathbf q}$. To first order in the perturbation,
\begin{equation}
\label{eq:appB_hamiltonian_matrix}
H
=
\frac12\mathbf p^{\mathsf T}\mathbf p
+
\frac12\mathbf q^{\mathsf T}\boldsymbol\Omega^2\mathbf q
-
\frac12\mathbf p^{\mathsf T}\mathbf C\mathbf p
+
\frac12\mathbf q^{\mathsf T}\mathbf K\mathbf q.
\end{equation}
For a single diagonal element, let $c_n(t)=C_{nn}(t)$ and $k_n(t)=K_{nn}(t)/\omega_{n0}^2$. A static perturbation gives
\begin{equation}
\label{eq:appB_static_frequency}
\omega_n'^2
=
\frac{\omega_{n0}^2[1+k_n]}
{1+c_n}
\simeq
\omega_{n0}^2(1+k_n-c_n),
\end{equation}
so $\delta\omega_n/\omega_{n0}=(k_n-c_n)/2$.

Using
\begin{equation}
\label{eq:appB_qp_operators}
q_n
=
\sqrt{\frac{\hbar}{2\omega_{n0}}}(a_n+a_n^\dagger),
\qquad
p_n
=-i\sqrt{\frac{\hbar\omega_{n0}}{2}}(a_n-a_n^\dagger),
\end{equation}
the diagonal pair part of the perturbation is
\begin{equation}
\label{eq:appB_pair_part}
\delta H_n^{\rm pair}(t)
=
\frac{\hbar\omega_{n0}}{4}
[c_n(t)+k_n(t)]
(a_n^2+a_n^{\dagger2}).
\end{equation}
For harmonic complex amplitudes $c_n(t)=\operatorname{Re}[\mathcal C_nh_0e^{-i\Omega t}]$ and $k_n(t)=\operatorname{Re}[\mathcal K_nh_0e^{-i\Omega t}]$, Eqs.~\eqref{eq:Romega_from_CK} and \eqref{eq:Rpair_definition} follow:
\begin{equation}
\label{eq:appB_two_responses}
R_n^\omega
=\frac{\mathcal K_n-\mathcal C_n}{2},
\qquad
R_n^{\rm pair}
=\frac{\mathcal K_n+\mathcal C_n}{2}.
\end{equation}
The equality of these two kernels is therefore an additional physical approximation, not an identity.

\section{Gaussian moment equations, positivity, and steady state}
\label{app:gaussian_dynamics}

For the Langevin equation
\begin{equation}
\label{eq:appC_langevin}
\dot a
=-\left(\frac\kappa2+i\delta\right)a
+2\lambda a^\dagger
+\sqrt{\kappa}\,a_{\rm in},
\end{equation}
with $\langle a_{\rm in}^\dagger(t)a_{\rm in}(t')\rangle=\bar n_{\rm b}\delta(t-t')$, the moments $N=\langle a^\dagger a\rangle$ and $M=\langle a^2\rangle$ obey
\begin{align}
\label{eq:appC_moments}
\dot N
&=-\kappa(N-\bar n_{\rm b})
+2\lambda M^*+2\lambda^*M,
\\
\dot M
&=-(\kappa+2i\delta)M
+4\lambda N+2\lambda.
\end{align}
The inhomogeneous term $2\lambda$ is the vacuum seed. At resonance, choose $\lambda>0$ and define $\mathcal G=4\lambda/\kappa$. The quadrature variances satisfy
\begin{align}
\label{eq:appC_variance_odes}
\dot V_{\rm amp}
&=-\kappa(1-\mathcal G)V_{\rm amp}
+\kappa(\bar n_{\rm b}+1/2),
\\
\dot V_{\rm sq}
&=-\kappa(1+\mathcal G)V_{\rm sq}
+\kappa(\bar n_{\rm b}+1/2).
\end{align}
For vacuum initial conditions, $V_{\rm amp}(0)=V_{\rm sq}(0)=1/2$, the solutions are Eq.~\eqref{eq:exact_transient_variances}. Since
\begin{equation}
\label{eq:appC_N_from_variances}
N(t)=\frac{V_{\rm amp}(t)+V_{\rm sq}(t)-1}{2},
\end{equation}
$N(t)$ is nonnegative for the exact solution. The Gaussian uncertainty product is also preserved:
\begin{equation}
\label{eq:appC_uncertainty}
V_{\rm amp}(t)V_{\rm sq}(t)\ge\frac14.
\end{equation}

In the stationary state,
\begin{equation}
\label{eq:appC_ss_variances}
V_{\rm amp}^{\rm ss}
=\frac{\bar n_{\rm b}+1/2}{1-\mathcal G},
\qquad
V_{\rm sq}^{\rm ss}
=\frac{\bar n_{\rm b}+1/2}{1+\mathcal G}.
\end{equation}
Using Eq.~\eqref{eq:appC_N_from_variances} gives
\begin{equation}
\label{eq:appC_Nss}
N^{\rm ss}
=\frac{\bar n_{\rm b}+\mathcal G^2/2}
{1-\mathcal G^2}.
\end{equation}
For $\bar n_{\rm b}=0$, this reduces to Eq.~\eqref{eq:vacuum_ss_photon_number}. The divergence as $\mathcal G\to1^-$ is the instability of the linear model; saturation requires nonlinear terms that are absent here.

\section{Anomalous output correlations and the pair spectrum}
\label{app:anomalous_output_correlations}
%===========================================================

We define the Fourier transform by
\begin{equation}
\hat a_{\rm out}(t)
=
\int_{-\infty}^{\infty}
\frac{d\omega}{2\pi}\,
e^{-i\omega t}\hat a_{\rm out}(\omega).
\label{eq:appB_fourier_definition}
\end{equation}
A monochromatic parametric modulation at angular frequency $\Omega_{\rm GW}$ creates photons in pairs whose frequencies obey
\begin{equation}
\omega+\omega'
=
\Omega_{\rm GW}.
\label{eq:appB_pair_energy_condition}
\end{equation}
To show how this constraint enters the output correlations, introduce slowly varying operators in the frame rotating at
$\Omega_{\rm GW}/2$:
\begin{equation}
\hat b_{\rm out}(t)
\equiv
e^{i\Omega_{\rm GW}t/2}\hat a_{\rm out}(t).
\label{eq:appB_rotating_operator}
\end{equation}
Below threshold, the driven-dissipative state is stationary in this rotating frame. Its anomalous two-time correlation therefore depends only on the time difference $\tau=t-t'$:
\begin{equation}
\left\langle
\hat b_{\rm out}(t)\hat b_{\rm out}(t')
\right\rangle
=
F(t-t').
\label{eq:appB_stationary_correlation}
\end{equation}
Transforming back to the laboratory frame gives
\begin{equation}
\left\langle
\hat a_{\rm out}(t)\hat a_{\rm out}(t')
\right\rangle
=
e^{-i\Omega_{\rm GW}(t+t')/2}
F(t-t').
\label{eq:appB_lab_correlation}
\end{equation}

The corresponding frequency-domain correlation is
\begin{align}
&\left\langle
\hat a_{\rm out}(\omega)
\hat a_{\rm out}(\omega')
\right\rangle
\nonumber\\
&\quad=
\int_{-\infty}^{\infty}dt
\int_{-\infty}^{\infty}dt'\,
e^{i\omega t+i\omega't'}
e^{-i\Omega_{\rm GW}(t+t')/2}
F(t-t').
\label{eq:appB_double_transform}
\end{align}
Using the variables
\begin{equation}
T=\frac{t+t'}{2},
\qquad
\tau=t-t',
\qquad
dt\,dt'=dT\,d\tau,
\label{eq:appB_time_variables}
\end{equation}
one obtains
\begin{align}
&\left\langle
\hat a_{\rm out}(\omega)
\hat a_{\rm out}(\omega')
\right\rangle
\nonumber\\
&\quad=
\int_{-\infty}^{\infty}dT\,
e^{i(\omega+\omega'-\Omega_{\rm GW})T}
\int_{-\infty}^{\infty}d\tau\,
e^{i(\omega-\omega')\tau/2}F(\tau).
\label{eq:appB_factorized_transform}
\end{align}
The integral over the mean time produces a Dirac delta function:
\begin{equation}
\left\langle
\hat a_{\rm out}(\omega)
\hat a_{\rm out}(\omega')
\right\rangle
=
2\pi\,
\delta\!\left(
\omega+\omega'-\Omega_{\rm GW}
\right)
M_{\rm out}(\omega),
\label{eq:appB_anomalous_correlator}
\end{equation}
where, on the support of the delta function,
\begin{equation}
M_{\rm out}(\omega)
\equiv
\int_{-\infty}^{\infty}d\tau\,
e^{i(\omega-\Omega_{\rm GW}/2)\tau}
F(\tau).
\label{eq:appB_anomalous_spectrum}
\end{equation}
Thus, the anomalous output spectrum correlates the two conjugate frequencies $\omega$ and $\Omega_{\rm GW}-\omega$.

The pair spectrum is the coefficient multiplying the distribution in Eq.~\eqref{eq:appB_anomalous_correlator}:
\begin{equation}
C_{\rm pair}(\omega)
\equiv
M_{\rm out}(\omega).
\label{eq:appB_pair_spectrum}
\end{equation}
A nonzero value of $C_{\rm pair}(\omega)$ is a phase-sensitive signature of coherent pair production. The ordinary output spectrum is defined by its stationary distribution,
\begin{equation}
\label{eq:appB_ordinary_spectrum}
\left\langle
\hat a_{\rm out}^{\dagger}(\omega)
\hat a_{\rm out}(\omega')
\right\rangle
=
2\pi\delta(\omega-\omega')S_{\rm out}(\omega).
\end{equation}
This spectrum measures occupation but does not, by itself, distinguish a phase-coherent pair source from an incoherent excess population.

At the degenerate parametric condition $\Omega_{\rm GW}\simeq2\omega_{n0}$, the correlated frequencies are centered symmetrically about the cavity resonance. In terms of the offset frequency $\nu$, they are
\begin{equation}
\omega
=
\omega_{n0}+\nu,
\qquad
\Omega_{\rm GW}-\omega
\simeq
\omega_{n0}-\nu.
\label{eq:appB_symmetric_frequencies}
\end{equation}
After the Dirac factor has been separated, the degenerate pair spectrum is
\begin{equation}
\label{eq:appB_degenerate_pair_correlation}
C_{\rm pair}(\nu)
\equiv
M_{\rm out}(\omega_{n0}+\nu).
\end{equation}
The spectral coefficient follows from the Langevin equation. Define rotating-frame Fourier operators by $a(t)=\int d\nu\,e^{-i\nu t}a[\nu]/(2\pi)$, choose the phase so that $\lambda_n$ is real, and write
\begin{align}
\label{eq:appB_DPA_frequency_coefficients}
d_{+,n}(\nu)
&=
\frac{\kappa}{2}-i(\nu-\delta_n),
\\
d_{-,n}(\nu)
&=
\frac{\kappa}{2}-i(\nu+\delta_n),
\\
\mathcal D_n(\nu)
&=
d_{+,n}(\nu)d_{-,n}(\nu)-4\lambda_n^2.
\end{align}
With $\xi[\nu]=\sqrt{\kappa_{\rm ext}}a_{\rm in}[\nu]+\sqrt{\kappa_{\rm int}}b_{\rm in}[\nu]$, the intracavity solution is
\begin{equation}
\label{eq:appB_intracavity_frequency_solution}
a[\nu]
=
\frac{
d_{-,n}(\nu)\xi[\nu]
+2\lambda_n\xi^{\dagger}[-\nu]
}
{\mathcal D_n(\nu)}.
\end{equation}
The input--output relation $a_{\rm out}=a_{\rm in}-\sqrt{\kappa_{\rm ext}}a$ gives
\begin{equation}
\label{eq:appB_output_scattering_matrix}
a_{\rm out}[\nu]
=
\sum_{j\in\{{\rm ext,int}\}}
\left[
S_j(\nu)c_j[\nu]
+
T_j(\nu)c_j^{\dagger}[-\nu]
\right],
\end{equation}
\begin{align}
\label{eq:appB_output_scattering_coefficients}
S_{\rm ext}(\nu)
&=
1-
\frac{\kappa_{\rm ext}d_{-,n}(\nu)}{\mathcal D_n(\nu)},
\\
T_{\rm ext}(\nu)
&=
-\frac{2\kappa_{\rm ext}\lambda_n}{\mathcal D_n(\nu)},
\\
S_{\rm int}(\nu)
&=
-\frac{\sqrt{\kappa_{\rm ext}\kappa_{\rm int}}d_{-,n}(\nu)}{\mathcal D_n(\nu)},
\\
T_{\rm int}(\nu)
&=
-\frac{2\sqrt{\kappa_{\rm ext}\kappa_{\rm int}}\lambda_n}{\mathcal D_n(\nu)}.
\end{align}
Here $c_{\rm ext}=a_{\rm in}$ and $c_{\rm int}=b_{\rm in}$. For vacuum inputs, $\langle c_j[\nu]c_k^{\dagger}[-\nu']\rangle=2\pi\delta_{jk}\delta(\nu+\nu')$, and the normally ordered and anomalous output spectra are
\begin{equation}
\label{eq:appB_explicit_output_spectrum}
S_{\rm out}^{(0)}(\nu)
=
\sum_j|T_j(\nu)|^2
=
\frac{4\kappa_{\rm ext}\kappa\lambda_n^2}
{|\mathcal D_n(\nu)|^2},
\end{equation}
\begin{align}
\label{eq:appB_explicit_anomalous_spectrum}
M_{\rm out}^{(0)}(\nu)
&=
\sum_jS_j(\nu)T_j(-\nu)
\nonumber\\
&=
-\frac{2\kappa_{\rm ext}\lambda_n}{\mathcal D_n(-\nu)}
+
\frac{
2\kappa_{\rm ext}\kappa\lambda_n d_{-,n}(\nu)
}
{\mathcal D_n(\nu)\mathcal D_n(-\nu)}.
\end{align}
The laboratory-frame coefficient is $M_{\rm out}(\omega)=M_{\rm out}^{(0)}(\omega-\Omega_{\rm GW}/2)$. Equations~\eqref{eq:appB_explicit_output_spectrum} and \eqref{eq:appB_explicit_anomalous_spectrum} show explicitly how the measurable spectrum depends on detuning and on both internal and external loss; finite-time or finite-bandwidth measurements are described by Eq.~\eqref{eq:finite_pair_correlation_detection}.

%===========================================================
\section{Output-quadrature squeezing in the presence of loss}
\label{app:output_quadrature_squeezing}
%===========================================================

In this appendix, $\hat a_{\rm out}$ denotes a normalized filtered output mode, such as the modes in Eq.~\eqref{eq:filtered_output_modes_detection}, rather than a distributional Fourier component. For a normalized output mode $\hat a_{\rm out}$, define the quadrature
operator
\begin{equation}
\hat X_{\theta}
=
\frac{1}{\sqrt{2}}
\left(
\hat a_{\rm out}e^{-i\theta}
+
\hat a_{\rm out}^{\dagger}e^{i\theta}
\right).
\label{eq:appC_quadrature_definition}
\end{equation}
For a zero-mean field, its variance is
\begin{align}
\left\langle
\Delta\hat X_{\theta}^{2}
\right\rangle
&=
\frac{1}{2}
\left\langle
\hat a_{\rm out}\hat a_{\rm out}^{\dagger}
+
\hat a_{\rm out}^{\dagger}\hat a_{\rm out}
\right\rangle
\nonumber\\
&\quad+
\operatorname{Re}
\left[
e^{-2i\theta}
\left\langle
\hat a_{\rm out}^{2}
\right\rangle
\right].
\label{eq:appC_general_variance}
\end{align}
Using $[\hat a_{\rm out},\hat a_{\rm out}^{\dagger}]=1$ and defining
\begin{equation}
N_{\rm out}
\equiv
\left\langle
\hat a_{\rm out}^{\dagger}\hat a_{\rm out}
\right\rangle,
\qquad
M_{\rm out}
\equiv
\left\langle
\hat a_{\rm out}^{2}
\right\rangle,
\label{eq:appC_output_moments}
\end{equation}
the variance becomes
\begin{equation}
\left\langle
\Delta\hat X_{\theta}^{2}
\right\rangle
=
\frac{1}{2}
+
N_{\rm out}
+
\operatorname{Re}
\left(
e^{-2i\theta}M_{\rm out}
\right).
\label{eq:appC_variance_moments}
\end{equation}
Writing $M_{\rm out}=|M_{\rm out}|e^{i\phi_M}$, the minimum is reached
for $2\theta=\phi_M+\pi$, giving
\begin{equation}
V_{\min}
\equiv
\min_{\theta}
\left\langle
\Delta\hat X_{\theta}^{2}
\right\rangle
=
\frac{1}{2}
+
N_{\rm out}
-
|M_{\rm out}|.
\label{eq:appC_minimum_variance}
\end{equation}
With the vacuum normalization
$V_{\rm vac}=1/2$, output squeezing occurs when
\begin{equation}
V_{\min}<\frac{1}{2},
\qquad\text{equivalently}\qquad
|M_{\rm out}|>N_{\rm out}.
\label{eq:appC_squeezing_condition}
\end{equation}

Internal loss and finite detection efficiency mix the ideal output field with independent environmental modes. If $\eta_{\rm tot}$ is the total power efficiency from the cavity output to the detector, the
measured mode may be written as
\begin{equation}
\hat a_{\rm meas}
=
\sqrt{\eta_{\rm tot}}\,
\hat a_{\rm out}
+
\sqrt{1-\eta_{\rm tot}}\,
\hat v,
\label{eq:appC_loss_beamsplitter}
\end{equation}
where $\hat v$ is an uncorrelated vacuum mode. The measured quadrature
is therefore
\begin{equation}
\hat X_{\theta}^{\rm meas}
=
\sqrt{\eta_{\rm tot}}\,
\hat X_{\theta}^{\rm out}
+
\sqrt{1-\eta_{\rm tot}}\,
\hat X_{\theta}^{v}.
\label{eq:appC_measured_quadrature}
\end{equation}
Because the two modes are uncorrelated and $\langle\Delta(\hat X_{\theta}^{v})^{2}\rangle=1/2$, the measured
variance is
\begin{equation}
V_{\theta}^{\rm meas}
=
\eta_{\rm tot}V_{\theta}^{\rm out}
+
\frac{1-\eta_{\rm tot}}{2}.
\label{eq:appC_loss_variance}
\end{equation}
Accordingly, the minimum observable variance is
\begin{equation}
V_{\min}^{\rm meas}
=
\frac{1}{2}
+
\eta_{\rm tot}
\left(
V_{\min}^{\rm out}-\frac{1}{2}
\right).
\label{eq:appC_measured_minimum_variance}
\end{equation}
Loss thus reduces the squeezing contrast relative to vacuum without changing its optimal quadrature angle.

For a cavity with external coupling rate $\kappa_{\rm ext}$ and internal loss rate $\kappa_{\rm int}$,
\begin{equation}
\kappa
=
\kappa_{\rm ext}
+
\kappa_{\rm int},
\label{eq:appC_total_linewidth}
\end{equation}
and the cavity escape efficiency is
\begin{equation}
\eta_{\rm esc}
=
\frac{\kappa_{\rm ext}}{\kappa}
=
\frac{\kappa_{\rm ext}}
{\kappa_{\rm ext}+\kappa_{\rm int}}.
\label{eq:appC_escape_efficiency}
\end{equation}
If the propagation and detector efficiency is $\eta_{\rm det}$, then
\begin{equation}
\eta_{\rm tot}
=
\eta_{\rm esc}\eta_{\rm det}.
\label{eq:appC_total_efficiency}
\end{equation}
More generally, if the environmental mode has thermal occupation $\bar n_{\rm env}$ rather than being in vacuum,
Eq.~\eqref{eq:appC_loss_variance} becomes
\begin{equation}
V_{\theta}^{\rm meas}
=
\eta_{\rm tot}V_{\theta}^{\rm out}
+
\left(1-\eta_{\rm tot}\right)
\left(
\bar n_{\rm env}+\frac{1}{2}
\right).
\label{eq:appC_thermal_loss_variance}
\end{equation}
This expression explicitly shows how internal loss, imperfect collection, and thermal noise degrade the observable squeezing.

%===========================================================
\section{Cavity sidebands and quasistatic flux-pumped conversion}
\label{app:cavity_sideband_derivation}

Consider a coherently driven cavity whose resonance frequency is
\begin{align}
\label{eq:appF_frequency_modulation}
\omega_n(t)
&=
\omega_{n0}
+
\delta\omega_{{\rm GW},n}
\cos(\Omega_{\rm GW}t-\theta_n^\omega),
\\
\delta\omega_{{\rm GW},n}
&=
\frac{\eta_n^\omega h_0\omega_{n0}}{2}.
\end{align}
Let $\Delta_d=\omega_d-\omega_{n0}$. In the drive frame,
\begin{equation}
\label{eq:appF_amplitude_equation}
\dot\alpha
=
\left[
 i\Delta_d-\frac\kappa2
-i\delta\omega_{{\rm GW},n}
\cos(\Omega_{\rm GW}t-\theta_n^\omega)
\right]\alpha
+
\sqrt{\kappa_{\rm ext}}\alpha_{\rm in}.
\end{equation}
With $\alpha=\alpha_0+\delta\alpha$ and
\begin{equation}
\label{eq:appF_alpha0}
\alpha_0
=
\frac{\sqrt{\kappa_{\rm ext}}\alpha_{\rm in}}
{\kappa/2-i\Delta_d},
\end{equation}
linearization gives
\begin{equation}
\label{eq:appF_sidebands}
\delta\alpha_\pm
=
-i\frac{\delta\omega_{{\rm GW},n}}{2}
 e^{\pm i\theta_n^\omega}
\alpha_0\chi_c(\pm\Omega_{\rm GW}),
\end{equation}
with $\chi_c(\nu)=[\kappa/2-i(\Delta_d+\nu)]^{-1}$. Therefore
\begin{equation}
\label{eq:appF_sideband_ratio}
\frac{|\delta\alpha_\pm|}{|\alpha_0|}
=
\frac{\eta_n^\omega h_0\omega_{n0}}{4}
|\chi_c(\pm\Omega_{\rm GW})|.
\end{equation}
For $\Delta_d=0$ and $\Omega_{\rm GW}\gg\kappa$,
\begin{equation}
\label{eq:appF_resolved_sideband}
\frac{|\delta\alpha_\pm|}{|\alpha_0|}
\simeq
\frac{\eta_n^\omega h_0\omega_{n0}}
{4\Omega_{\rm GW}}.
\end{equation}
At $\Omega_{\rm GW}\simeq2\omega_{n0}$, this should not be interpreted as a pair of resonantly enhanced sidebands of the same mode; the degenerate process is described by the pair Hamiltonian instead.

For the flux-pumped conversion channel, define $\mathcal F_n(\Phi)=\omega_n(\Phi)R_n^\omega(\Phi)$ and $\Phi(t)=\Phi_b+\Phi_p\cos(\Omega_pt+\varphi_p)$. Expansion gives Eq.~\eqref{eq:flux_response_cross_term}. The difference-frequency amplitude is
\begin{equation}
\label{eq:appF_IF_amplitude}
\delta\omega_{\rm IF}
=
\frac{\Phi_ph_0}{2}
|\mathcal F_n'(\Phi_b)|,
\qquad
\Omega_{\rm IF}=|\Omega_{\rm GW}-\Omega_p|.
\end{equation}
The corresponding probe sidebands obey
\begin{equation}
\label{eq:appF_IF_sidebands}
\frac{|\delta\alpha_{{\rm IF},\pm}|}{|\alpha_0|}
=
\frac{\delta\omega_{\rm IF}}{2}
|\chi_c(\pm\Omega_{\rm IF})|.
\end{equation}
This derivation uses the flux dependence of the calibrated frequency-response function and introduces no GW-induced SQUID flux. It applies only in the quasistatic and off-resonant regime specified by Eq.~\eqref{eq:off_resonant_flux_pump_condition}. When the pump approaches $2\omega_{n0}$, the direct coupling in Eq.~\eqref{eq:direct_flux_pump_pair_coupling} and the corresponding Floquet scattering problem must be included.

%\nocite{*}

\bibliography{apssamp}% Produces the bibliography via BibTeX.

\end{document}